\documentclass[pra,letterpaper,twocolumn,superscriptaddress,floatfix]{revtex4}
\usepackage{graphicx,psfrag,amsmath,amssymb,amsfonts,bbm,latexsym,color,dcolumn,bm}

\begin{document}
	
\title{Dynamics of Thermalization in classical three-dimensional many-body systems: \\ non-Maxwellian distributions and the role of anisotropic trapping}

\author{Roberto Onofrio}

\affiliation{\mbox{Department of Physics and Astronomy, Dartmouth College, 6127 Wilder Laboratory, 
		Hanover, NH 03755, USA}}

\author{Bala Sundaram}

\affiliation{\mbox{Department of Physics, University of Massachusetts, Boston, MA 02125, USA}}

\begin{abstract}
The physics of interacting baths is of interest in a variety of contexts, ranging from ultracold atoms to ionized gases. 
While many features can already be captured in a one-dimensional model, others are specific to the situation of 
two or three-dimensional systems. In this paper, we focus on three-dimensional features of a model for 
thermalization between two baths that we have explored in earlier publications. In contrast to the one-dimensional situation we show that, enroute to thermalization, deviations from the Maxwell-Boltzmann energy distributions more akin to $\kappa$-distributions are observed. In three dimensions, these can be seen in changes at the peak of the distribution, which is precluded by the expected 
monotonic behavior of the energy distributions in one dimension.
Furthermore, we consider thermalization for anisotropic trapping and explore its role in the 
onset of anomalous scaling with respect to the particle number. 
These results may be pertinent to the understanding of the interplay between turbulent 
behavior and nonlinearities for a variety of many-body classical systems.
\end{abstract}

\maketitle

\section{Introduction}

The transfer of energy  between many-body systems is critical to understanding dissipative processes and equilibration
dynamics. In a conservative system setting, based on Hamiltonian dynamics, the Caldeira-Leggett model
\cite{Magalinskii,Ullersma1,Ullersma2,Ullersma3,Ullersma4,Caldeira1,Caldeira2,Caldeira3}
captures most of the physics via the linear coupling of a subsystem, composed of a test particle or a set of
harmonic oscillators, to another subsystem made up of an infinite number of harmonic oscillators having a continuum of
frequencies.

Motivated by the need to understand thermalization processes in systems composed of a finite and
discrete spectrum of frequencies - as seen for trapped systems in atomic and plasma physics - 
we considered thermalization in a model where both the range and the strength of the interaction
appeared in a nonlinear generalization of the Caldeira-Leggett model~\cite{OnoSun,JauOnoSun1}.
Later, the model was also used to explore the interplay of nonlinearities arising either or both
from interaction and confining potentials~\cite{JauOnoSun2}.
As a byproduct of the nonlinearity, we reported in~\cite{JauOnoSun1} anomalous scaling for the average
total interaction energy  as a function of the number of particles, for two equally balanced baths, as thermalization was approached.
The exponent associated with the power-law scaling was found to be rather close to that associated with Kolmogorov scaling
in turbulent fluid mixing. We further explored the dependence of this anomalous scaling on the strength of the inter-particle interaction~\cite{JauOnoSun3}, showing that the anomalous scaling is seen for an intermediate range of values of the interaction strength. 
While our initial focus was on one-dimensional systems, several phenomena  persisted in higher dimensions, differing only in the specific value of the anomalous scaling exponent. A perturbative analytic approach allowed us to predict the values of the anomalous scaling exponents, or at least to establish bounds on their values.  More recently, we explored deviations from the Maxwell-Boltzmann (MB) energy distributions during thermalization, which were due to large interspecies coupling and interaction range. Further, we also
identified the generation of the convolution of two MB energy distributions at distinct temperatures, bimodal MB (bMB) energy distributions  \cite{Bala10}, of considerable interest in plasma physics in the context of enhancing fusion reactivity 
 \cite{Majumdar,Onofrio,Xie1,Xie2,Kong,Squarer,Ourabah}. However, we were unable to identify the presence of a $\kappa$ distribution expected for long range interactions \cite{Konishi,Ruffo}.

\begin{figure*}[t]
\includegraphics[width=0.49\textwidth, clip=true]{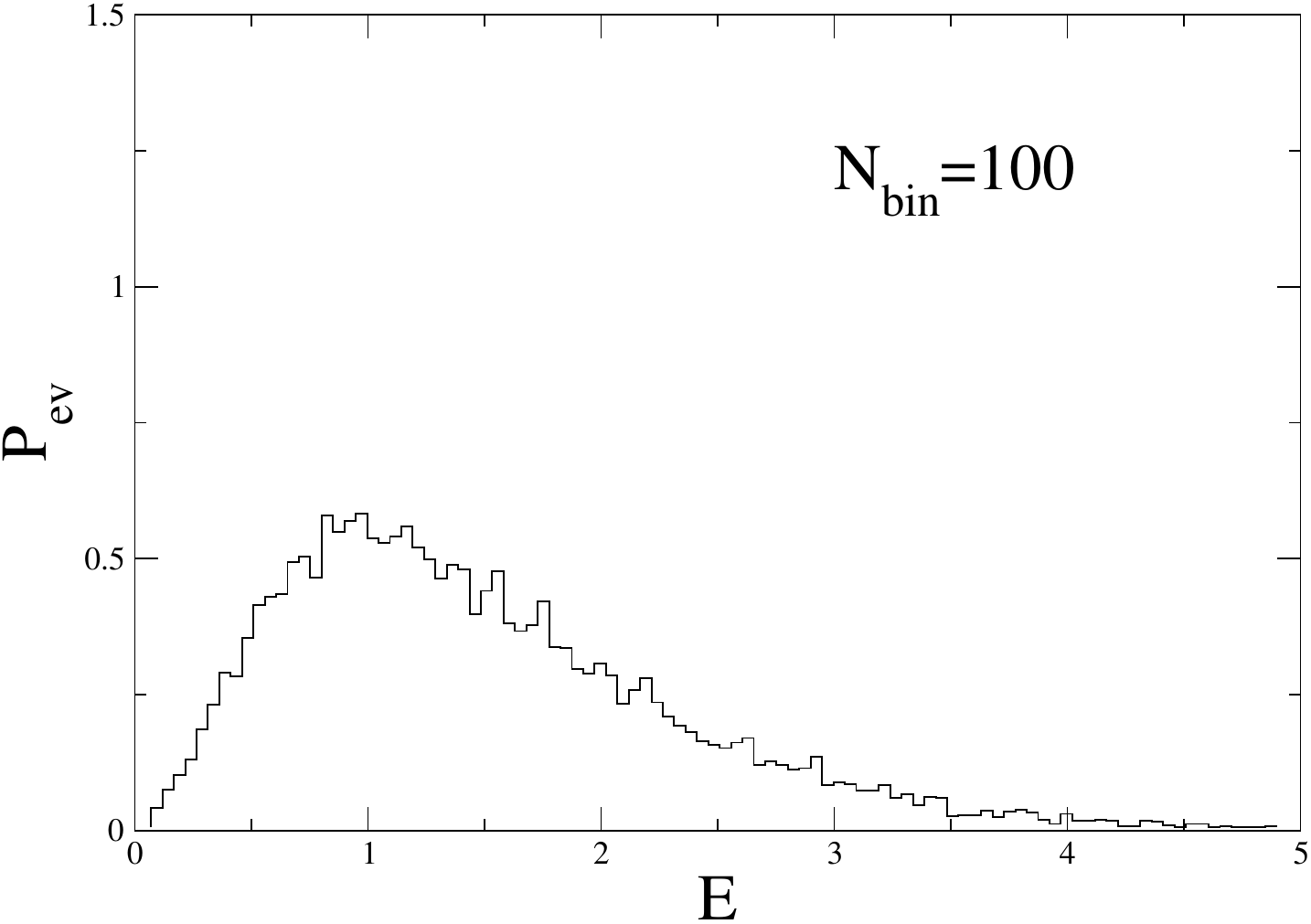}
\includegraphics[width=0.49\textwidth, clip=true]{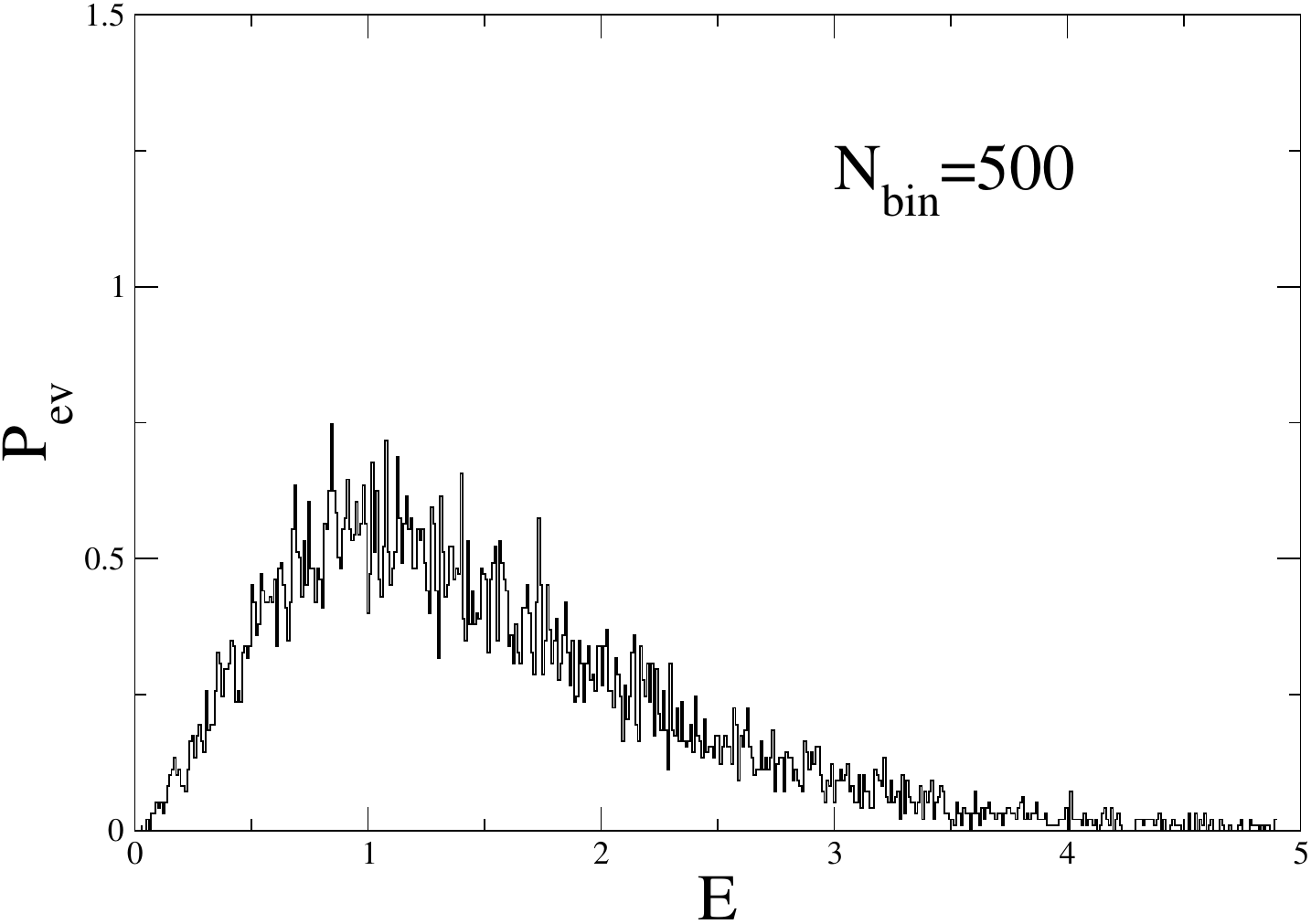}
\includegraphics[width=0.49\textwidth, clip=true]{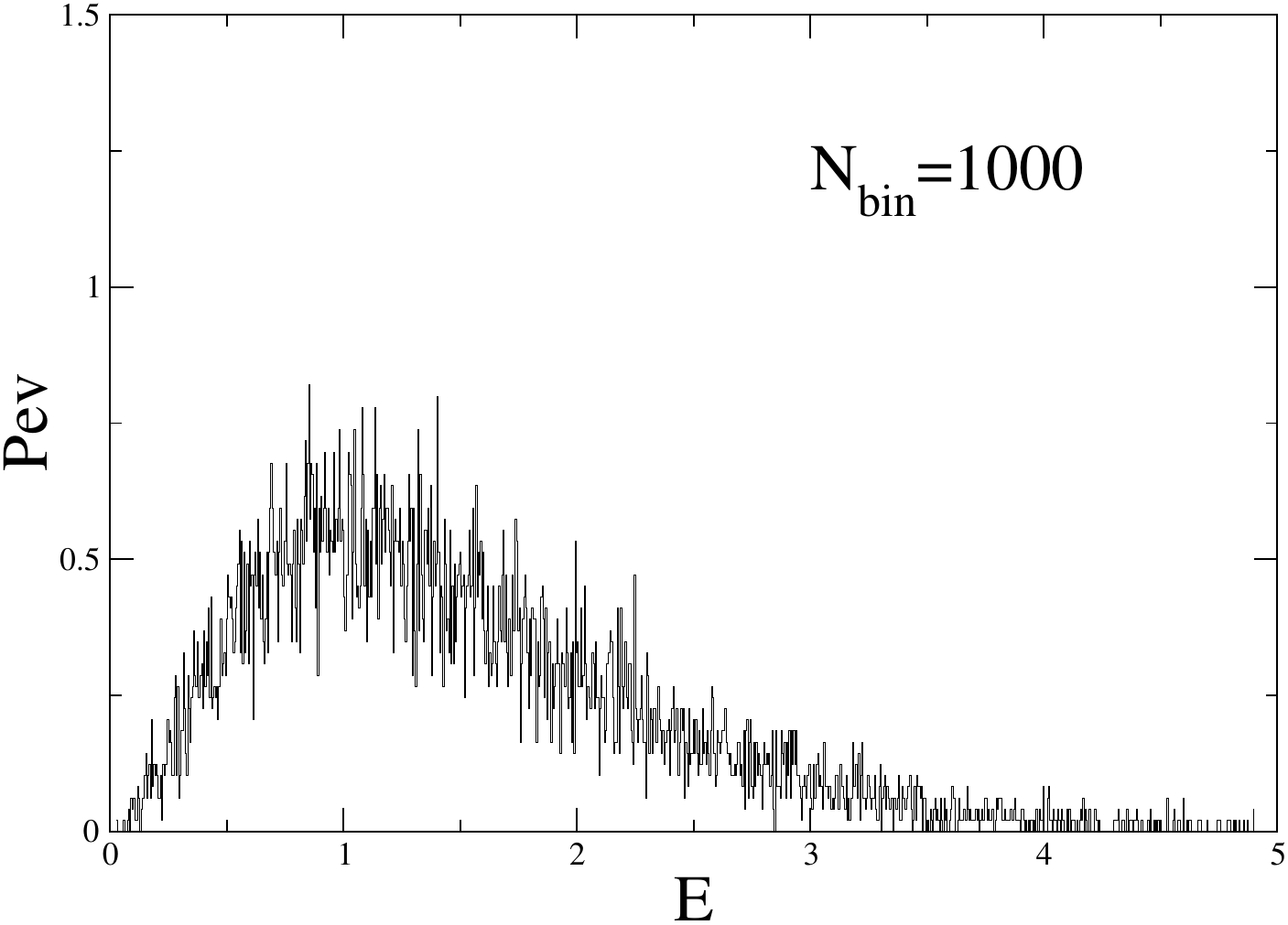}
\includegraphics[width=0.49\textwidth, clip=true]{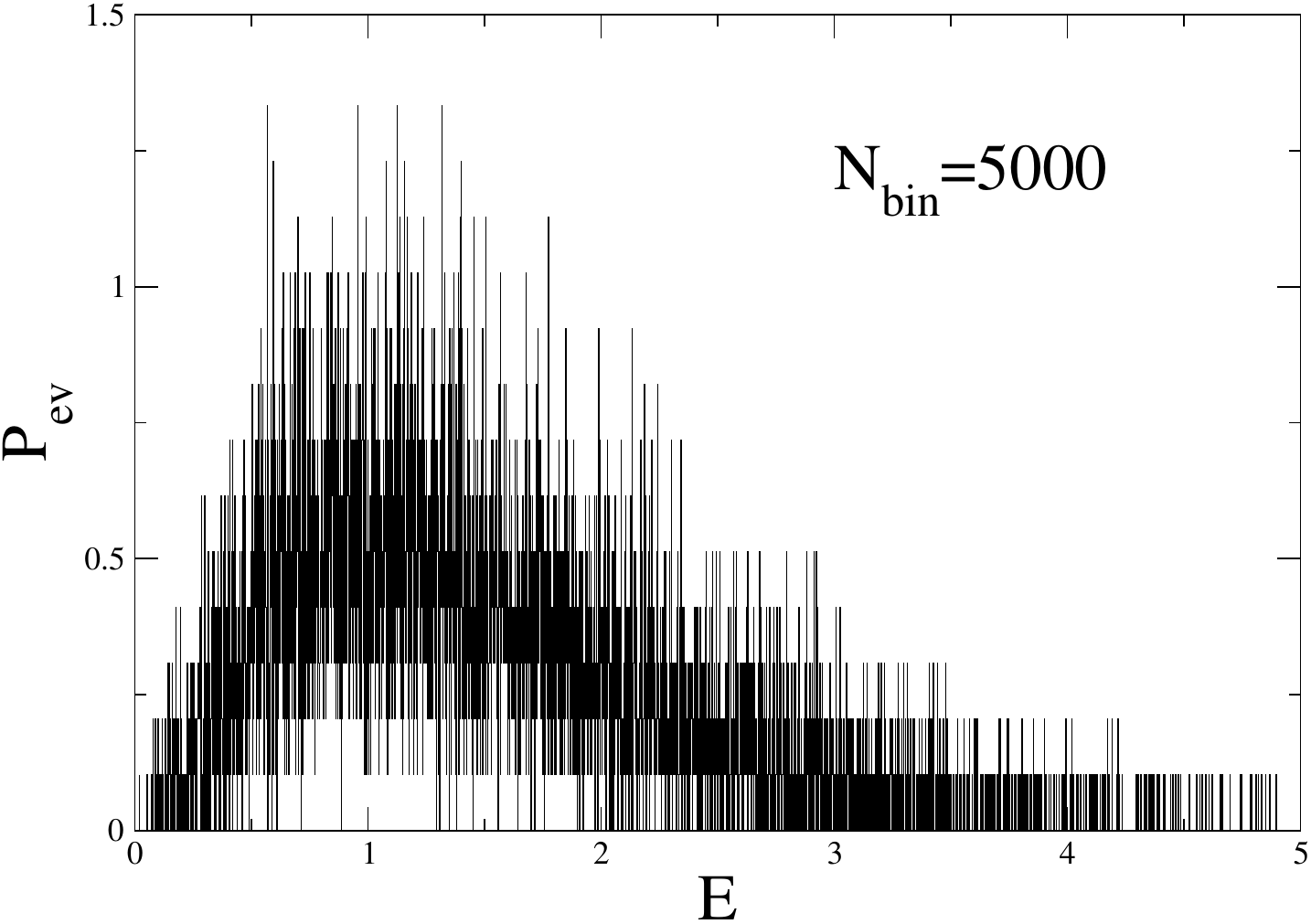}
\caption{Histograms of the same Maxwell-Boltzmann energy distribution generated by a sample of $10^4$ 
particles with inverse temperature $\beta=2.0$ and four choices of energy binning with $N_{\mathrm{bin}}=100, 500, 1000, 5000$,  
$E_{\mathrm{min}}=0.06931$ and $E_{\mathrm{max}}=4.89855$, such that the energy bin is $(E_{\mathrm{max}}-E_{\mathrm{min}})/N_{\mathrm{bin}}$.}
\label{Fig1}
\end{figure*}

\begin{figure*}[t]
\includegraphics[width=0.48\textwidth, clip=true]{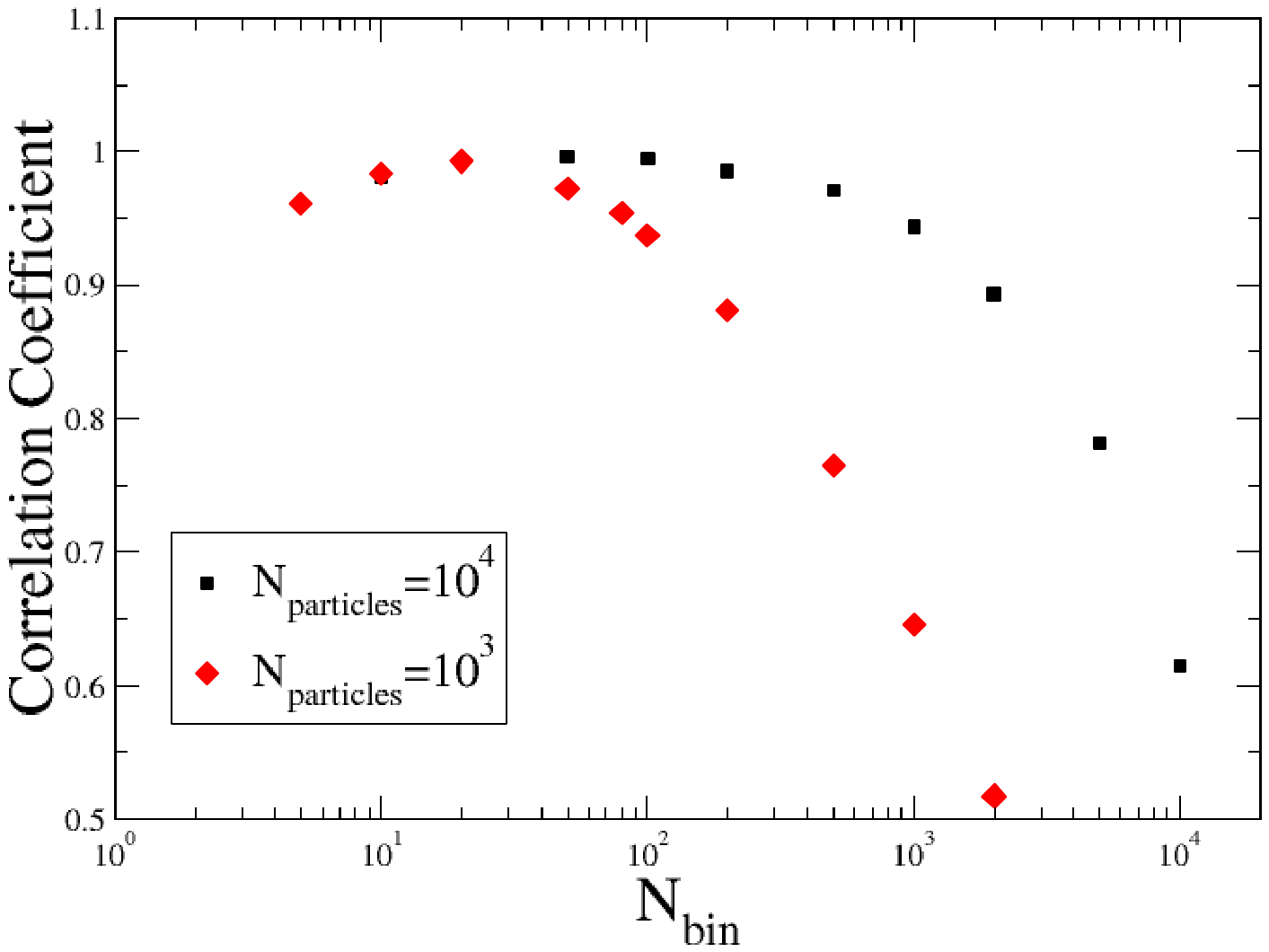}
\includegraphics[width=0.48\textwidth, clip=true]{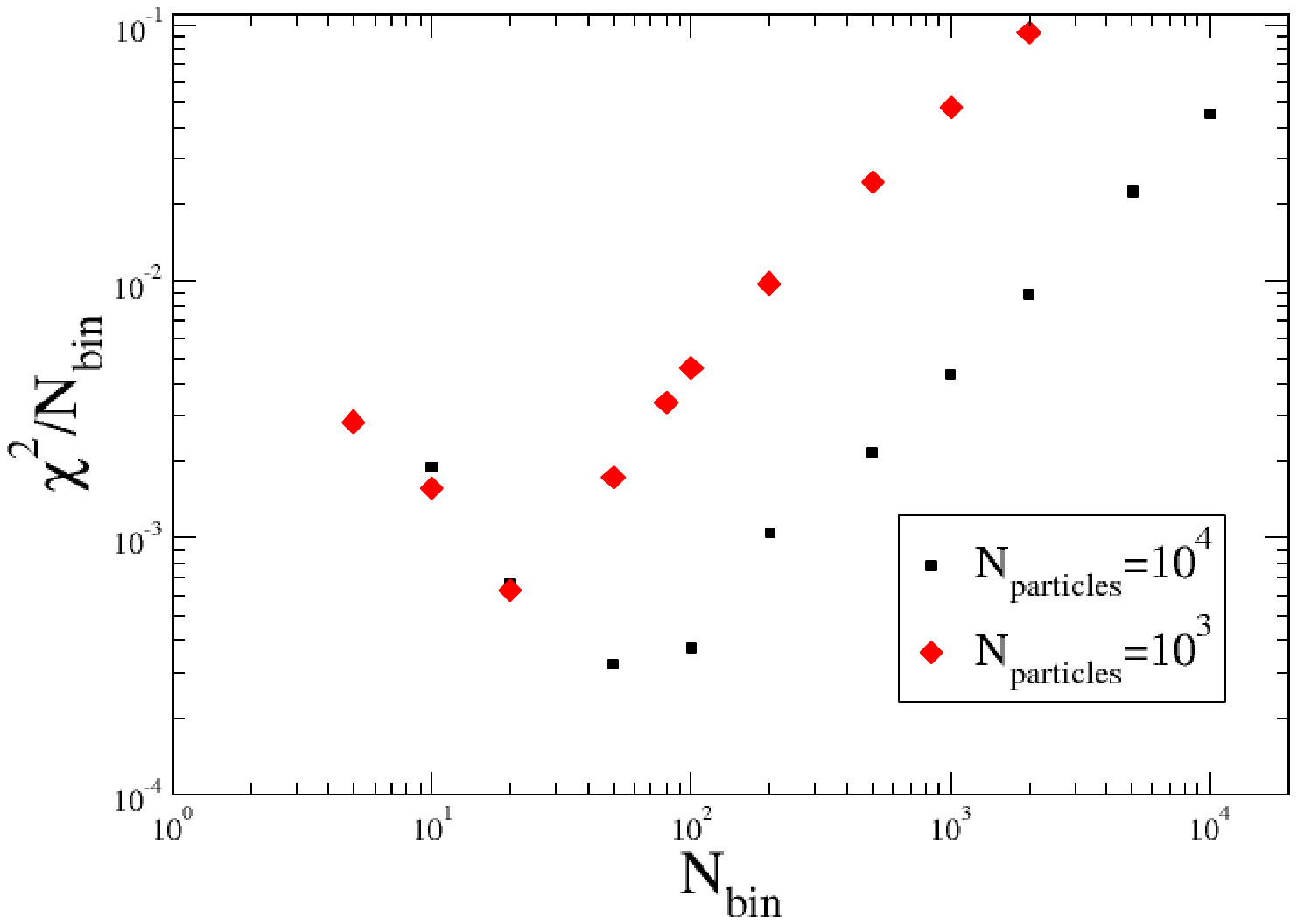}
\includegraphics[width=0.48\textwidth, clip=true]{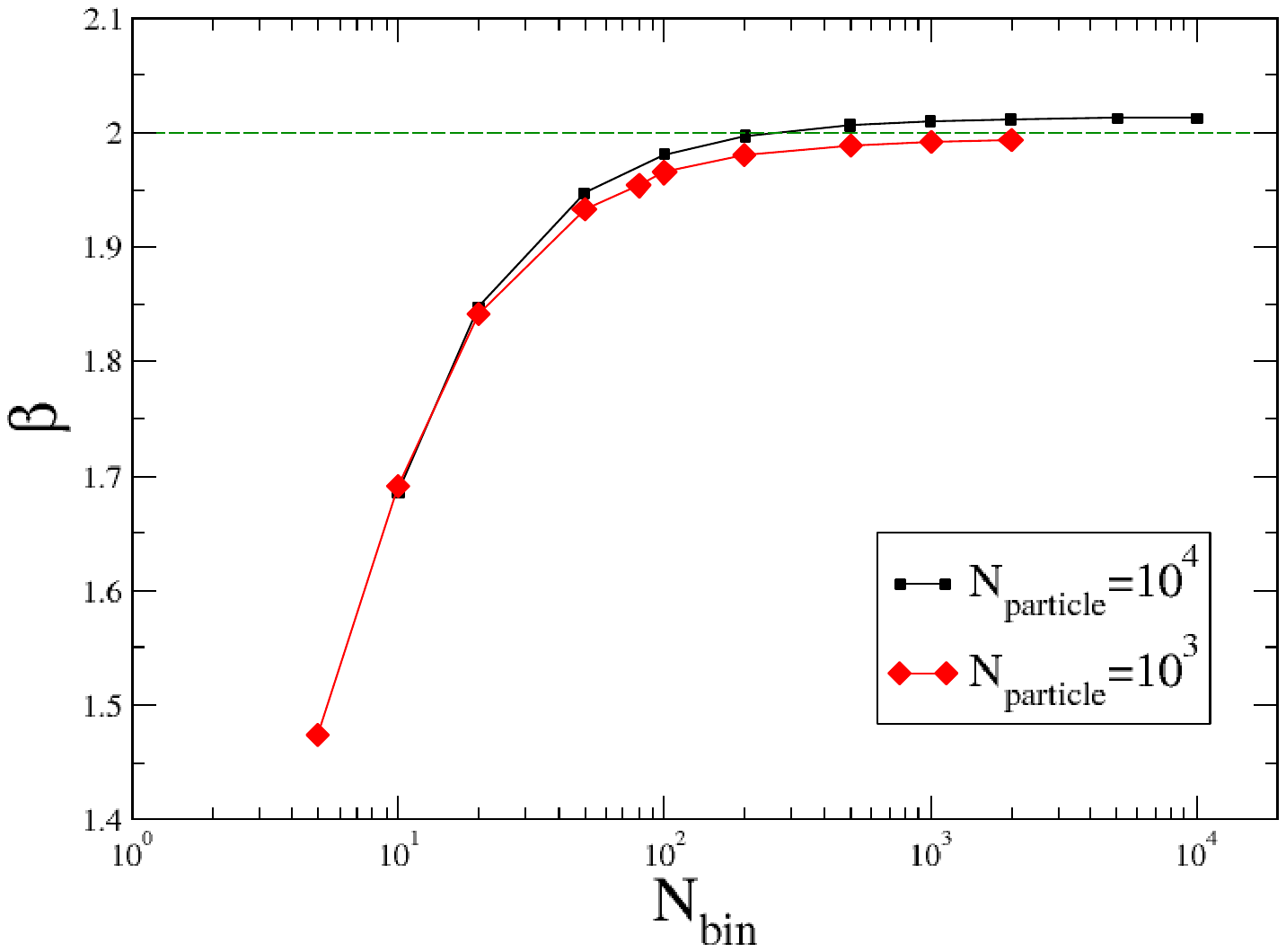}
\includegraphics[width=0.48\textwidth, clip=true]{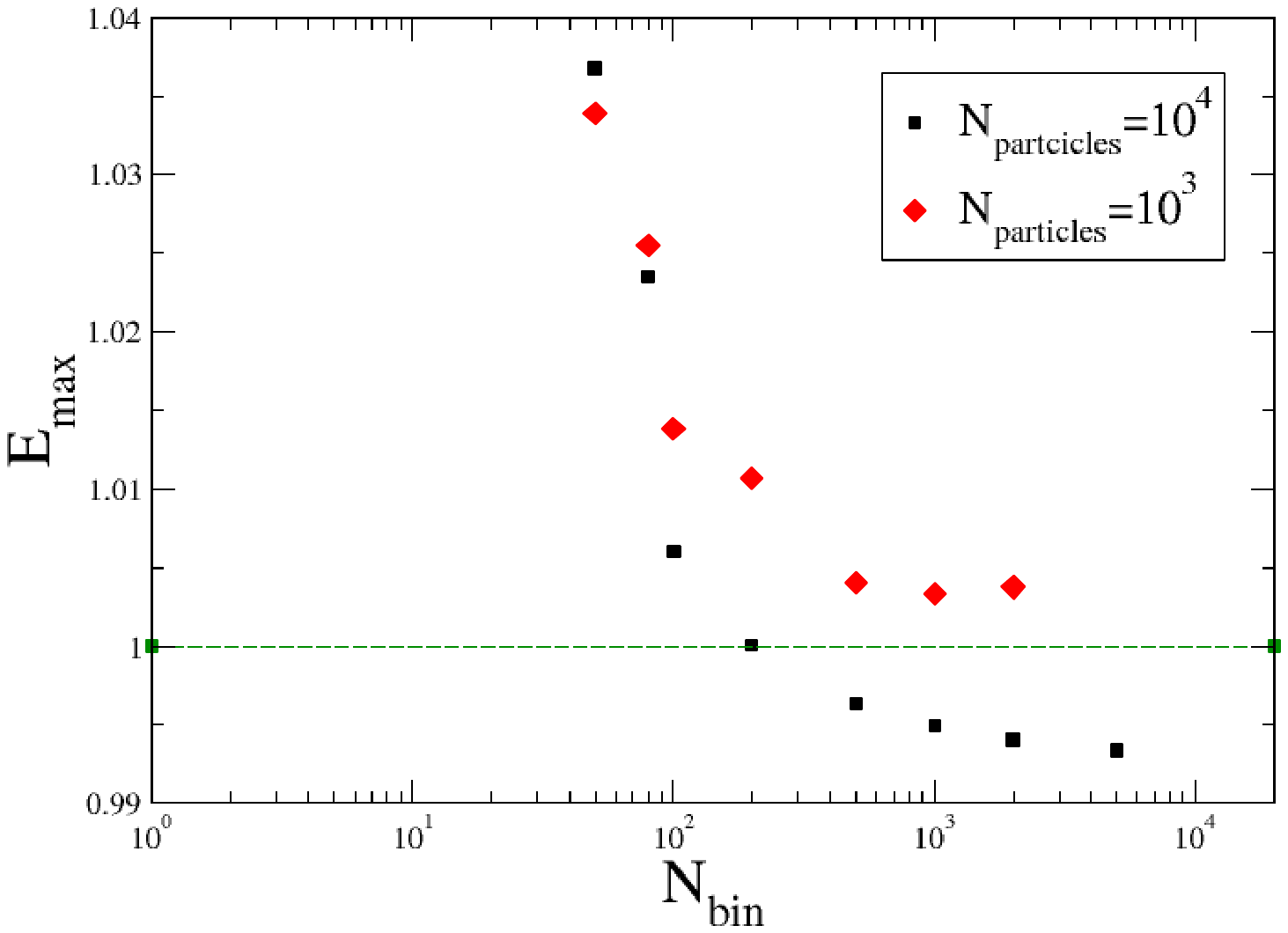}
\caption{Analysis of the histograms in Fig.~\ref{Fig1}, and additional cases, for determining the best fit with respect to $N_{\mathrm{bin}}$. 
The correlation coefficient (top-left) shows a maximum around $N_{\mathrm{bin}} \simeq 50$, confirmed by the reduced 
$\chi^2$ indicator (top-right). The curve of the inverse temperature shows instead a value closer to the one of the parent 
population for $N_{\mathrm{bin}} \simeq 300$ (bottom-left), and this trend is confirmed in the behavior of the fitting curves 
around the peak value (bottom-right), with the case of $N_{\mathrm{bin}}=500$ being peaked closer to the 
expected value of $E_{\mathrm{max}}=1$ than the case of $N_{\mathrm{bin}}=100$. The dashed lines in the lower panels
reflect the expected values for a MB energy distribution with inverse temperature $\beta=2$ (left plot)and the energy at which the energy distribution reaches the  maximum, at $E_{\mathrm{max}}=2/\beta=1$ (right plot).}
\label{Fig2}
\end{figure*}

Here, we continue the exploration of the model by extending our analysis to peculiar features that are more clearly
manifest only in higher dimensions. In Section II, we start with a preliminary analysis of the sensitivity of the peak of the energy 
distribution for a three dimensional system of harmonic oscillators, determining the appropriate system size
necessary for the numerics presented later. In Section III, we consider thermalization dynamics between two species, with the goal of 
demonstrating the appearance of non-MB distributions, specifically $\kappa$-distributions enroute to thermalization.
In Section IV we discuss the case of thermalization for anisotropic trapping, which allows us to interpolate between the genuine 
one-dimensional and three-dimensional cases. We continue with an analysis of the anomalous scaling, a reflection of non-extensivity, 
under more general conditions. This helps establish the analogy between the anomalous exponents that we observe with those seen in magnetohydrodynamics. 
General considerations on the applicability of our model findings to concrete physical contexts are discussed  in the concluding section.

\section{Determination of the peak of the energy distributions}

As discussed in a recent contribution~\cite{Bala10}, it is difficult to demonstrate, with 
numerical simulations, the appearance of $\kappa$-distributed energies 
during thermalization in one-dimensional many-body systems initially prepared with MB energy distributions. A previous numerical study focused on the 
dynamics of systems initially prepared  with $\kappa$ energy distributions that relaxed to MB distributions with time~\cite{BagchiTsallis}.

The most prominent effect of $\kappa$-distributions is the presence of a high-energy 
tail with power-law scaling, rather than the exponential scaling  seen for MB energy distributions. This in principle is a 
remarkable difference, though, for a finite sample of particles, the statistical error in the tails is too large to clearly distinguish
the two energy distributions. However, the situation changes on going to higher dimensions, as here the energy distribution, due to the different density of states, has a peak. 
at intermediate energies which provides a clear distinction between the two distributions.
 For example, in three dimensions, the probability density for 
the MB energy distribution  of an ensemble of harmonic oscillators can be written as

 \begin{equation}
p(E)= \frac{1}{2} \beta^3 E^2 e^{-\beta E}, 
\label{Eqp}
 \end{equation} 
which has a maximum at 

\begin{equation}
E_{\mathrm{max}}=\frac{2}{\beta},
\label{MBEmax}
\end{equation}
corresponding to a peak probability density equal to 
\begin{equation}
p_{\mathrm{max}}=p(E_{\mathrm{max}})=2 e^{-2} \beta.
\label{MBpmax}
\end{equation}
These relationships hold rigorously for a bath of noninteracting particles under harmonic trapping but, 
given the strength of the interactions in our situation, these
should be taken as reliable approximations in the weak coupling limit, as discussed later. 
\begin{figure*}[t]
\includegraphics[width=0.49\textwidth, clip=true]{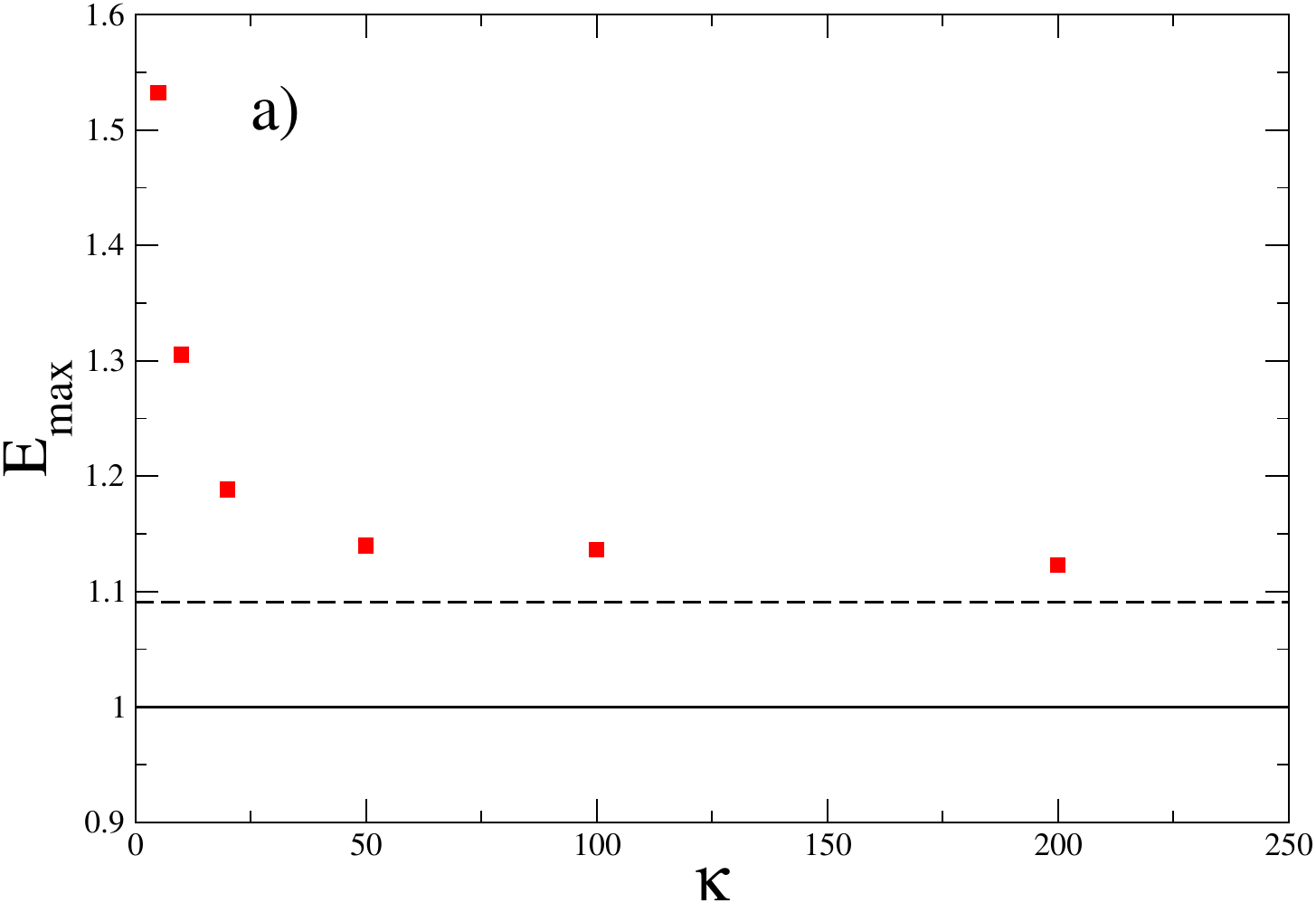}
\includegraphics[width=0.49\textwidth, clip=true]{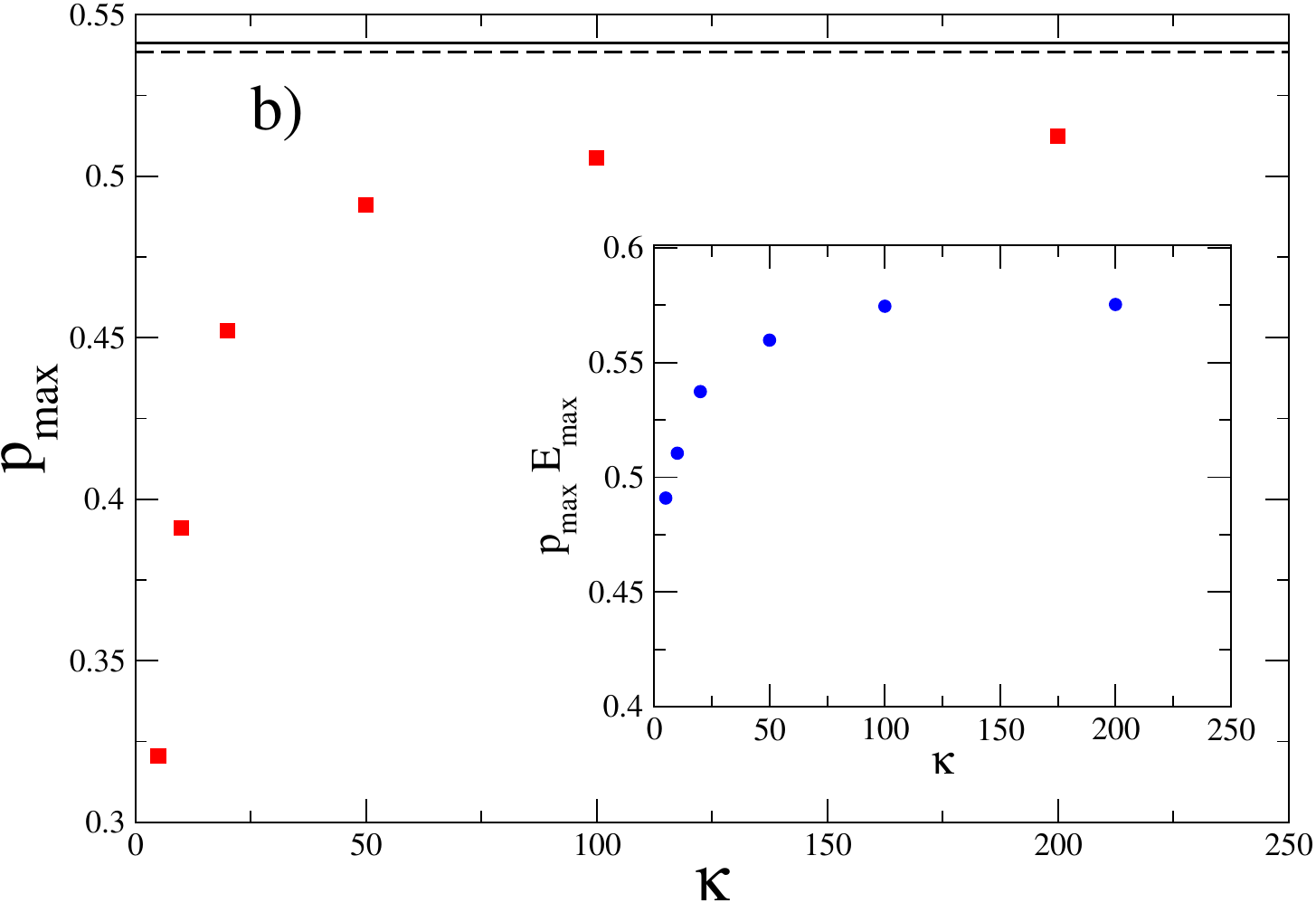}
\caption{Dependence on $\kappa$ of the energy for which the $\kappa$-distribution has a maximum in the probability density (a), and corresponding 
value of the probability density (b) determined from a sample of $10^4$ particles, with their energies partitioned into $10^2$ bins, assuming $\beta=2$. In each plot, the continuous line shows the theoretical expectation from a MB distribution for 
 the same inverse temperature, as from Eqs. (\ref{MBEmax}) and (\ref{MBpmax}), respectively. 
The dashed line shows the corresponding realization obtained with a fit to an energy distribution, generated from 
properly weighted random numbers and the same nominal inverse temperature, with an MB distribution, 
for the same sample of $10^4$ particles. In the inset in (b) we plot the product $p_{\mathrm{max}} E_{\mathrm{max}}$ versus $\kappa$.  The corresponding MB case with same number of particles and bins  has a value 
$(p_{\mathrm{max}} E_{\mathrm{max}})_{MB} = 0.5872$, against an expected value of $4 e^{-2} \simeq 0.54134$.
All values of quantities with physical dimensions here and in the following figures are in arbitrary units.}
\label{Fig3}
\end{figure*}
The presence of a peak is advantageous as we can look for deviations at the peak of the distribution, 
where statistical errors are smaller. However, we first need to assess the sensitivity of the peak position, given the finite
number of particles and the associated
choice of binning when arriving at the histogram of the energy distribution. In practice, we do not assess the probability density, but rather
a coarse-grained quantity, the number of events with energy in between $E$ and $E+\Delta E$, related to the probability density as 
\begin{equation}
P_{\mathrm{ev}}^{(j)}= \int_{E_j}^{E_j+\Delta E} p(E) dE.
\end{equation}
$\Delta E=(E_{\mathrm{max}}-E_{\mathrm{min}})/N_{\mathrm{bins}}$, where $E_{\mathrm{max}}$ and $E_{\mathrm{min}}$ 
are the maximum and minimum energies obtained in a equal-time configuration, and $N_{\mathrm{bin}}$ is the number of 
bins used for the distribution. In order to optimize the sensitivity to energy shifts, we have generated a nominally Maxwell-Boltzmann 
energy distribution for $10^4$ particles in each bath,  with  well-defined inverse temperatures. Then, for varying choices of the 
number of bins, we have constructed the energy histogram and fitted with a MB distribution. As seen in Fig.~\ref{Fig1}, the 
choice of the binning affects the energy distribution. As expected, for a small number of bins (panel (a)), the statistical 
fluctuations in the number of particles is less but the energy resolution is low. On increasing the number of bins, the energy
resolution improves at the cost of increased statistical 
fluctuations. This is particularly evident in panel (d) in which 5,000 bins are chosen, 
{\it i.e.} an extreme case with only an average of two events each bin. To optimize this choice we consider, in Fig.~\ref{Fig2},
two indicators of the fidelity of the fitting procedure, the correlation coefficient and the reduced $\chi^2$. Using these, we are able to determine the impact on the two relevant parameters of the expected distribution, the inverse temperature $\beta$ and the related 
maximum of the energy distribution, $E_{\mathrm{max}}$.

As seen from Fig.~\ref{Fig2},  the peak of the energy distribution depends on the binning, and 
the optimal binning seems close to $N_{\mathrm{bin}} \simeq 200$. This is at variance with trends from the correlation 
coefficient and the $\chi^2$ analyses, likely because the latter are figures expressing the agreement over the entire 
range of the fit. By repeating the analysis for baths consisting of $10^3$ oscillators we find, not surprisingly, that the curves 
for the various indicators are shifted towards lower precision for the same number of bins. It is worth noting that the curves 
for the measurement of the temperature are not significantly affected by this tenfold decrease in the bath populations. Instead, the 
dependence of the energy for which the peak of the distribution occurs is significantly shifted and the optimal value is achieved for 
a larger value of the number of bins. With reference to the bottom-right panel in Fig.~\ref{Fig2}, $E_{\mathrm{max}}=1.0$ is achieved for 
$N_{\mathrm{bin}}=200$ in the case of $10^4$ particles in each bath. Instead, the closest approximant in the case of $10^3$ particles 
in each bath (but differing from $E_{\mathrm{max}}$ by about 4 $\%$) is achieved at $N_{\mathrm{bin}}=10^3$. 

\begin{figure}[t]
\includegraphics[width=0.49\textwidth, clip=true]{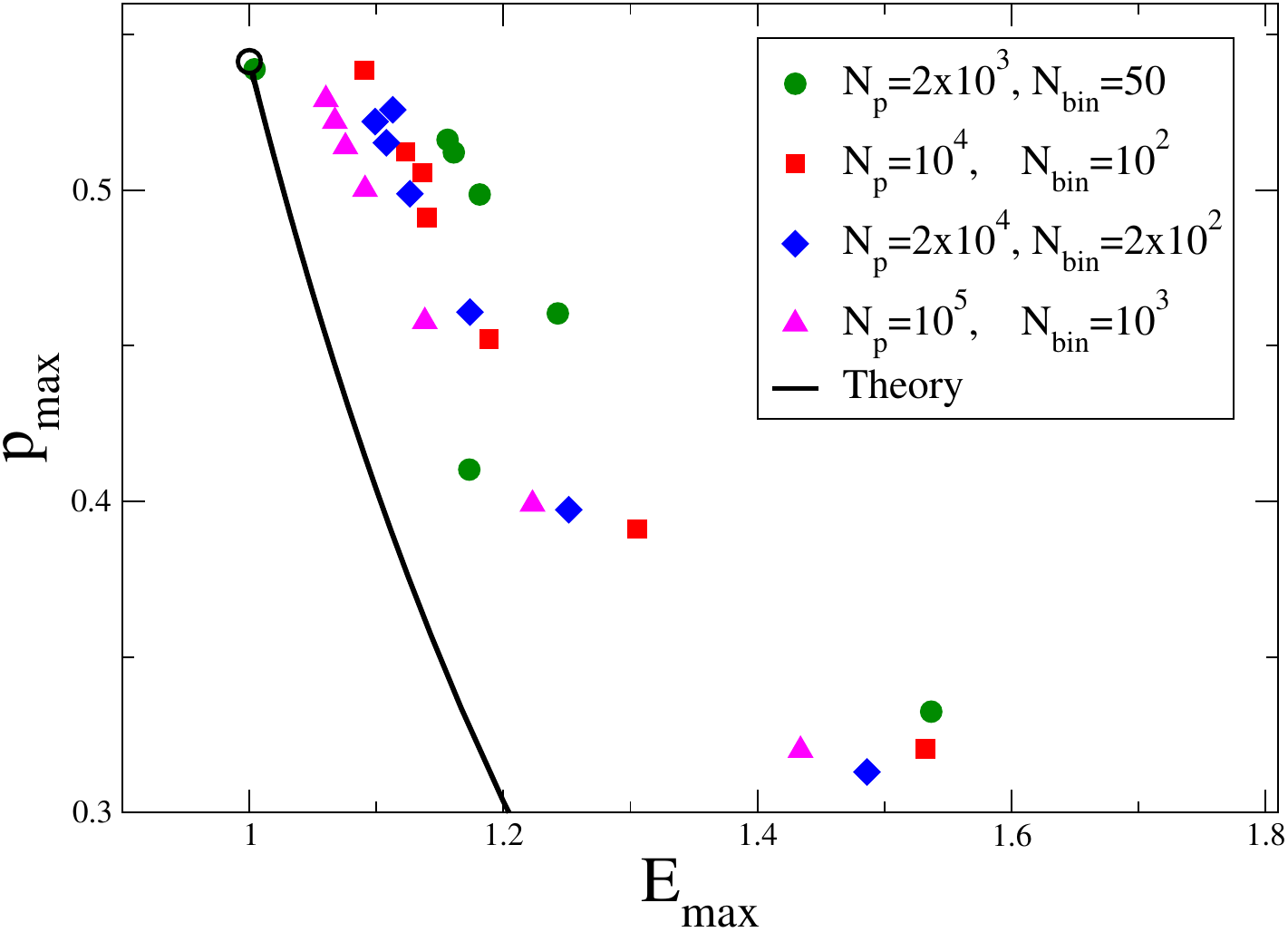}
\caption{Plot in the $E_{\mathrm{max}}-p_{\mathrm{max}}$ plane, with points obtained for various values of $\kappa$ ($\kappa=5, 10, 20, 50, 100, 200$ from 
bottom right, plus the point for the corresponding Maxwell-Boltzmann sample on top left) with different sample populations and binnings. 
The continuous curve is obtained from Eqs. \ref{KappaEmax} and \ref{Kappapmax}, and the case of the Maxwell-Boltzmann 
probability density is highlighted as a empty circle on the upper-left of the curve.}
\label{Fig4}
\end{figure}

A similar analysis can be repeated for $\kappa$-distributions where again, for a noninteracting bath embedded in a harmonic confinement potential, the 
probability density \cite{LivadiotisPotential} is given by
 
\begin{equation}
p(E)=\frac{1}{2} \beta^3 E^2\frac{(\kappa-2)(\kappa-1)}{\kappa^2} \left(1+\frac{\beta E}{\kappa}\right)^{-\kappa-1}.
\end{equation}
There  are now two parameters, the inverse temperature $\beta$ of the equivalent MB energy distribution that results
for $\kappa \to \infty$, and the $\kappa$-parameter ($\kappa >2$). The maximum value of the distribution occurs at 
\begin{equation}
E_{\mathrm{max}}=\frac{2}{\beta} \frac{\kappa}{\kappa-1}
\label{KappaEmax}
\end{equation} 
with corresponding  peak probability density
\begin{equation}
p_{\mathrm{max}}= 2 \frac{(\kappa-2)(\kappa-1)}{(\kappa+1)^{\kappa+1}} \beta.
\label{Kappapmax}
\end{equation}
In Fig. \ref{Fig3} we show the values of the energy at which the $\kappa$ probability distribution peaks and the related probability
as a function of $\kappa$, for a simulation involving $10^4$ particles and a partition of their energies into $10^2$ bins.
These values are obtained by focusing on a region around the peak of the distribution, rather than over the entire range. The distribution is 
filtered using a cutoff, selecting only those bins having probabilities higher than 60 $\%$ of $p_{\mathrm{max}}$. This selection is then 
fitted with a parabola of the form 
\begin{equation}
p(E)=p_{\mathrm{max}}+c (E-E_{\mathrm{max}})^2,
\end{equation}
where $c$ is the curvature at the peak value. In order to assess the sensitivity of the result to the sample population, we 
show the values of $E_{\mathrm{max}}$ and $p_{\mathrm{max}}$ obtained from the corresponding MB distribution 
with the same number of particles  and bins, as well as the predictions from Eqs. (\ref{MBEmax}-\ref{MBpmax}). It should be noted that, for
the MB distribution, the peak of the distribution is quite sensitive to the sample size.  For our parameter choices, the deviation is around 10 $\%$ from 
the predicted value from Eq. \ref{MBEmax}. By contrast,  the probability density is far more robust, with a deviation of less that 1 $\%$, a 
fact attributable to the curvature of the probability distribution. A more comprehensive assessment of the sensitivity to the sample size 
is obtained by plotting the two quantities shown in Fig.~\ref{Fig3} against each other. The result is shown in Fig.~\ref{Fig4} which considers several choices of particle
number and binning as well as the theoretical curve. As expected, increasing the sample population improves the agreement with theory, 
as seen for $10^5$ particles. Moreover, at smaller sample populations there is more scattering in the points, most notably 
in the case of $2 \times 10^3$ particles. This is a direct consequence of the lower statistics.
We also verified that the systematic shift towards the right 
with respect to the theoretical curve is due to the asymmetry of the probability energy distribution around the peak, 
with a slower decrease at energies higher than $E_{\mathrm{max}}$. This can be corrected by choosing a higher cut-off though the corresponding
cost of smaller statistics of selected events leads to larger scattering of the data which are, on average, closer to the curve.

\section{Thermalization dynamics: Transient emergence of kappa-distributions} 

Having  established the conditions under which MB and $\kappa$ energy distributions can be distinguished by contrasting the peaks of the distributions, 
we can now consider the dynamics of thermalization
between two ensembles of particles which are harmonically trapped.  We consider the  Hamiltonian already introduced in \cite{OnoSun,JauOnoSun1,JauOnoSun2}

\begin{eqnarray}
H&=&\sum_{m=1}^{N_A} \left(\frac{P_m^2}{2m_A}+\frac{1}{2} m_A \omega_A^2 Q_m^2 \right)+ \nonumber \\
& & \sum_{n=1}^{N_B} \left(\frac{p_n^2}{2m_B}+\frac{1}{2} m_B \omega_B^2 q_n^2 \right)+ \nonumber \\
& & \gamma \sum_{m=1}^{N_A} \sum_{n=1}^{N_B} 
\exp{\left[-\frac{\left(Q_m-q_n\right)^2}{\lambda^2}\right]},
\label{Hamilton}
\end{eqnarray}
where $(Q_m,P_m)$ and $(q_n,p_n)$ are the positions and momenta of each particle of the
two species $A$ and $B$, respectively, where all variables lie in a generic D-dimensional space.
The interspecies term is governed by two parameters, with $\gamma$ being the strength
and $\lambda$ the range of the interaction. The equations of motion corresponding to the Hamiltonian 
in Eq.~\ref{Hamilton} can be numerically integrated to machine precision. Runs with $10^3$ particles in 
each bath requires a few hours of computer time, allowing for a compromise between the size of the 
bath and the possibility to  quickly explore various features of the dynamics. 
Indeed, this Hamiltonian facilitates the study of a variety of situations, including balanced ($N_A=N_B$) 
and unbalanced mixtures, attractive ($\gamma<0$) and repulsive ($\gamma>0$) interactions, as well as  
long-range ($\lambda \rightarrow \infty$) and short-range interactions. 
In particular, for a completely unbalanced mixture (for instance $N_A=1$ and $N_B \rightarrow \infty$),
small $\gamma$ and large interaction range, the model mimics, in the classical limit, the genuine
Caldeira-Leggett approach used to model dissipation in open systems. We begin by considering
isotropic trapping, before moving to the more general anisotropic setting.

\begin{figure}[t]
\includegraphics[width=0.49\textwidth, clip=true]{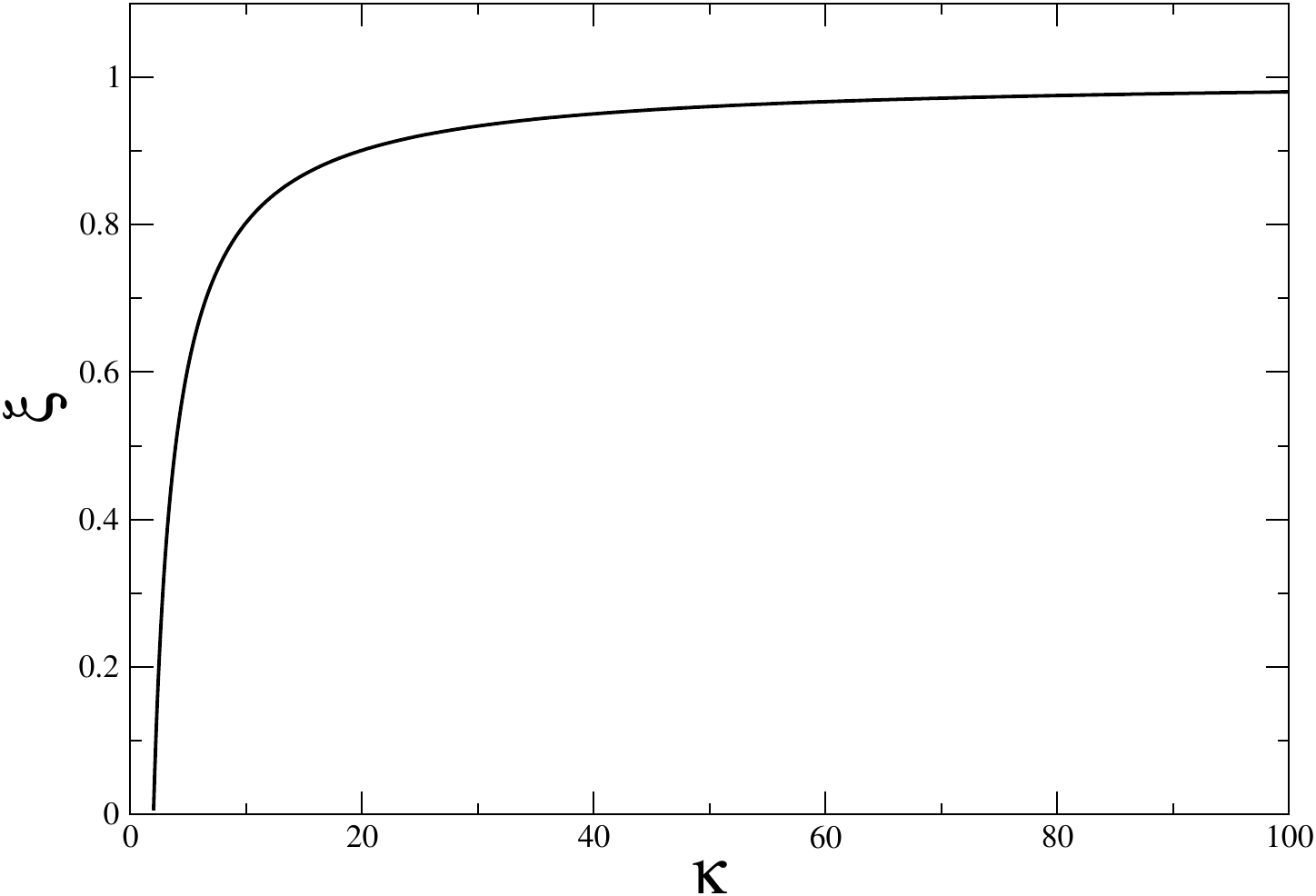}
\caption{Dependence of the $\xi$ parameter defined in Eq. \ref{XiDef} on $\kappa$.}
\label{Fig5}
\end{figure}

\begin{figure*}[t]
\includegraphics[width=0.49\textwidth, clip=true]{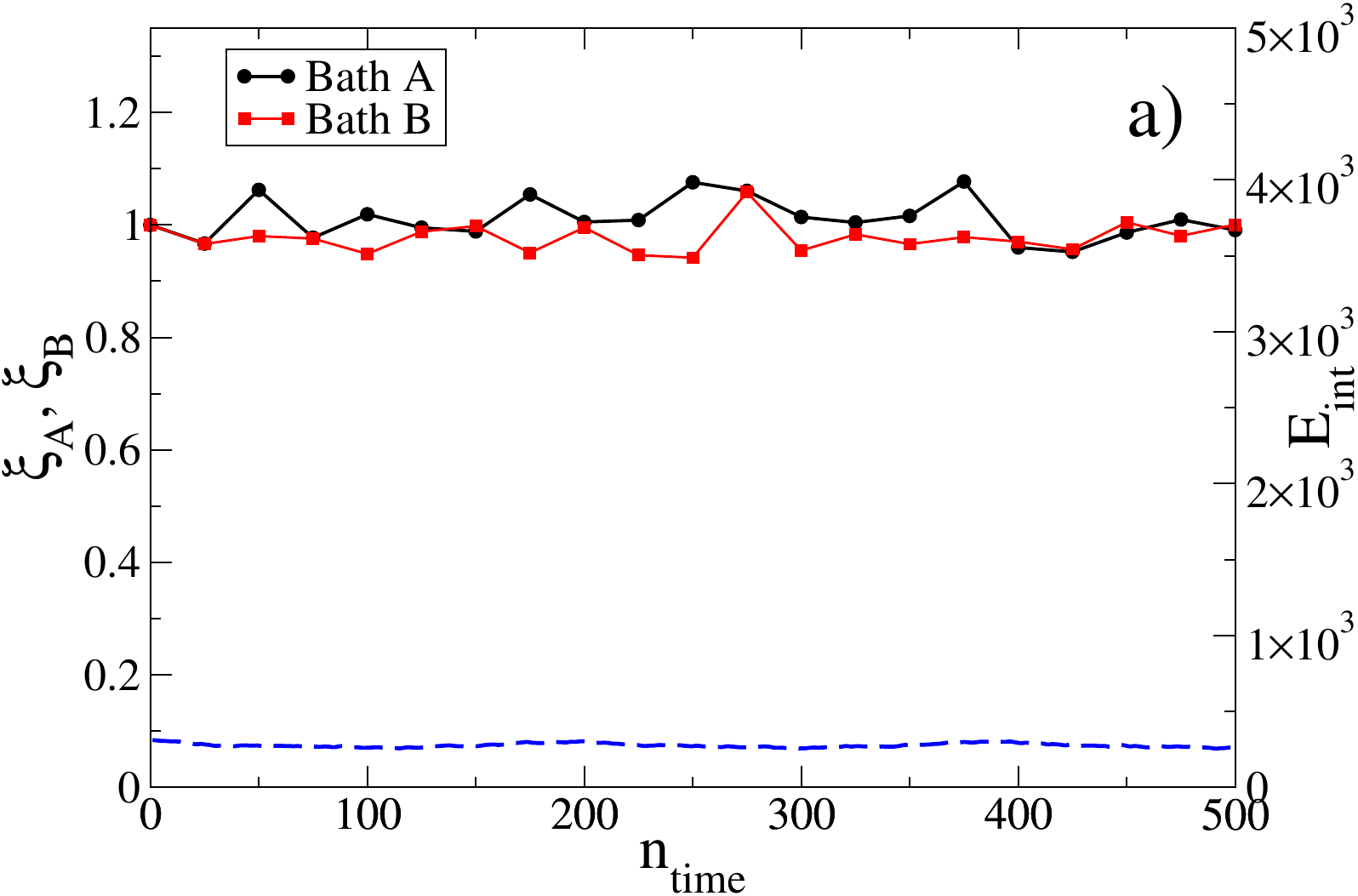}
\includegraphics[width=0.49\textwidth, clip=true]{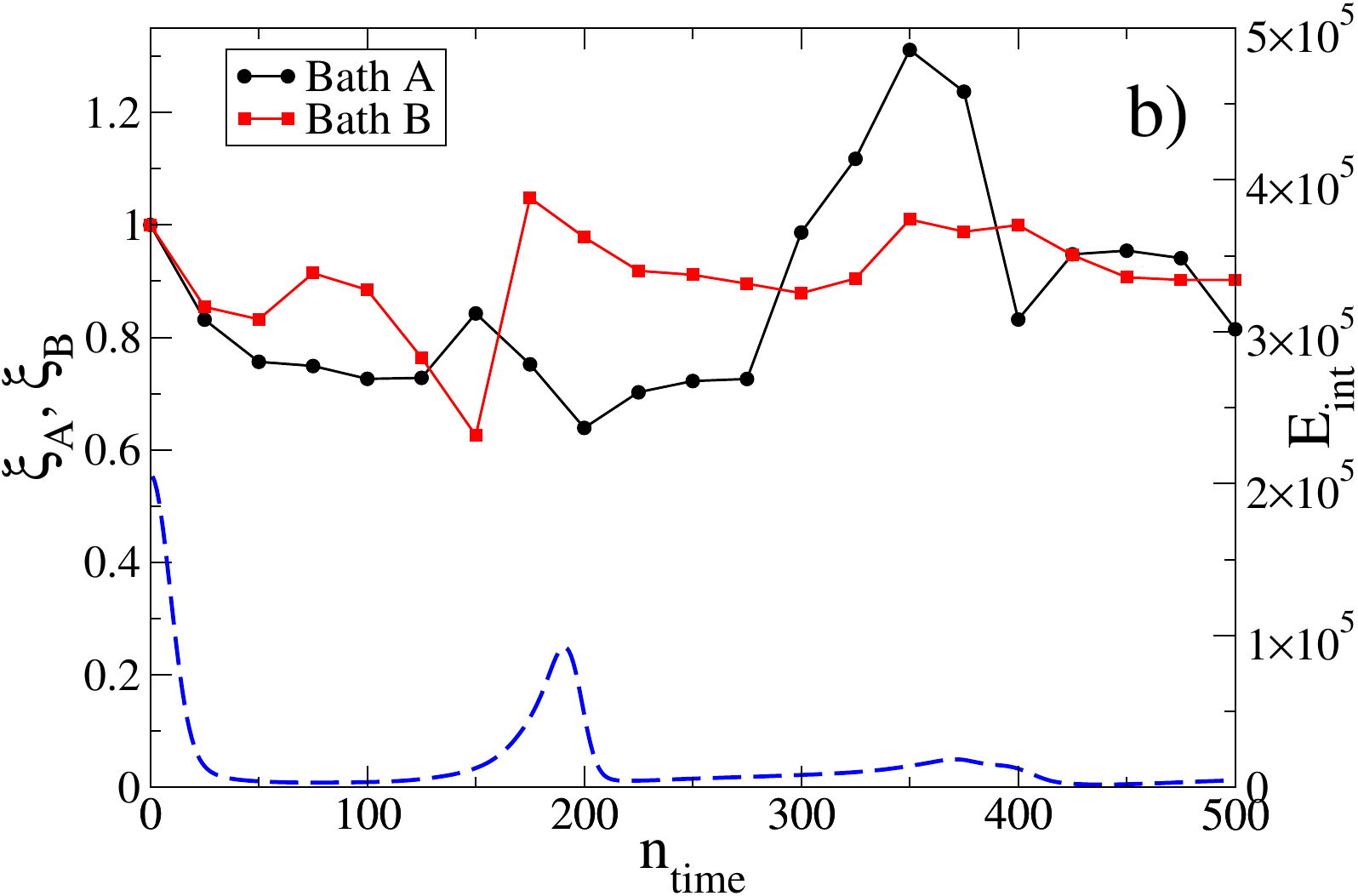}
\caption{Dependence of the $\xi$ on time (measured in time steps) for two baths made of 5,000 particles each with initial inverse temperatures of $\beta_A=2$ and $\beta_B=0.2$, $\gamma=0.1$, and two values of $\lambda=10^{-1}$ (a) and $1$ (b), with $m_A=m_B$ and isotropic trapping with $\omega_A=144/89, \omega_B=1$ . The analysis is made by partitioning the particle energies into 200 bins, with a threshold for the determination of the interpolating parabola equal to 0.50 of the maximum value in the data set. In the left plot the energy distribution remains MB at all time, while in the right 
plot the colder bath (A) experience an initial strong deviation from a MB energy distribution reaching a value of $\xi \simeq 0.6$ after 200 time steps, then increasing to reach $\simeq 1.3$ after 350 times steps, before resettling to a constant value compatible with unity. The time dependence of the interaction energy is also shown, lower dashed curves, with the vertical scales on the right side of each plot. Notice that there is a factor 100 in the values of $E_\mathrm{int}$ for the two plots. In plot (b) revivals of the interaction energy are evident at about 180 time steps and, less prominent but still visible, at about 370 time steps, in close proximity of the largest deviations from MB distributions for the two baths.}
\label{Fig6}
\end{figure*}

We first focus on the dynamics of thermalization, more specifically the possibility that, during the 
earlier stages of thermalization, the two baths evolve with non-MB features, most notably developing $\kappa$-distributed 
energies. Obviously the two baths will evolve and change their temperatures during the interaction process, and this would appear to 
complicate the analysis described in the previous section. However, in the limit of weak interactions, we note that the product $p_{\mathrm{max}} E_{\mathrm{max}}$ is independent on temperature. This dimensionless product is

\begin{equation}
(p_{\mathrm{max}} E_{\mathrm{max}})_{MB} = 4 e^{-2} \simeq 0.54134,
\end{equation}
for a Maxwell-Boltzmann energy distribution, and 

\begin{equation}
(p_{\mathrm{max}} E_{\mathrm{max}})_\kappa= \frac{4(\kappa-2)\kappa(\kappa-1)^{\kappa-1}}{(\kappa+1)^{\kappa+1}},
\end{equation} 
for a $\kappa$ energy distribution. Therefore, it is natural to introduce a parameter $\xi$ defined as
\begin{equation}
\xi(\kappa)=\frac{(p_{\mathrm{max}} E_{\mathrm{max}})_{\kappa}}{(p_{\mathrm{max}} E_{\mathrm{max}})_{MB}}=
\frac{(\kappa-2)\kappa(\kappa-1)^{\kappa-1}}{(\kappa+1)^{\kappa+1}}e^2.
\label{XiDef}
\end{equation}

This ratio versus $\kappa$ is shown in Fig. \ref{Fig5}, where clearly $\xi \to 1$ in the $\kappa \to \infty$ limit, to coincide
with a MB distribution. Note also that $\xi \simeq 0.9$ for $\kappa \simeq 20$, and $\xi \simeq 0.8$ for $\kappa \simeq 10$.
The associated inference is that as long as errors arising from violations of the weakly interacting assumption, the finite sample sizes and the binning and fitting procedures are kept within 10 $\%$, a $\kappa$ distribution of value $\kappa \leq 20$ is still discernible from the corresponding MB distribution at the same temperature. Unfortunately, the sensitivity to changes in $\kappa$ decreases for low values where deviations from MB would be most pronounced.

 We have analyzed the time evolution of energy distributions in the case of 5,000 particles in each bath, 
evaluating the corresponding $\xi$ by creating histograms for the energy distribution with 200 bins. The same value of $\gamma$ was used for two different $\lambda$ values. The first
$\lambda$ was closer to short range interactions, as in Fig.~\ref{Fig6}a, and the other a tenfold increase in $\lambda$ that, for the choice of the other parameters, represents long-range interactions, in Fig.~\ref{Fig6}b. While in the first case the value of $\xi$ is compatible with unity at all times, in the second case we notice strong deviations, more specifically values of $\xi$ close to 0.6 after 150 time steps for bath B, and after 200 steps for bath A. Furthermore, we observe values of $\xi$ of order 1.2-1.3 at later times, around 350 time steps, for bath A only. In general, we expect that the colder bath to be more 
sensitive to the transient and the exchange of interaction energy, which is consistent with our numerical results.
These results are quite robust with respect to various choices of parameters in the data analysis. In particular, we contrasted the results on changing the 
probability cutoff value from $50 \%$ to $65 \%$ of the peak value. Setting the threshold too high decreases the number of selected points for the parabolic fit, while a low threshold gives rise to systematic shifts in the determination of $E_{\mathrm{peak}}$ due to the asymmetric form of the  collected probability distribution. For the two choices above, we obtained plots for $\xi$ versus time steps differing by less than 5 $\%$. Also, the fact that $\xi$ recovers higher values at later times after the drops at a value of about 0.6 indicates that the chosen value of $\gamma$ is low enough to make the expected deviations from the MB negligible on the considered timescale.

\begin{figure*}[t]
\includegraphics[width=0.49\textwidth, clip=true]{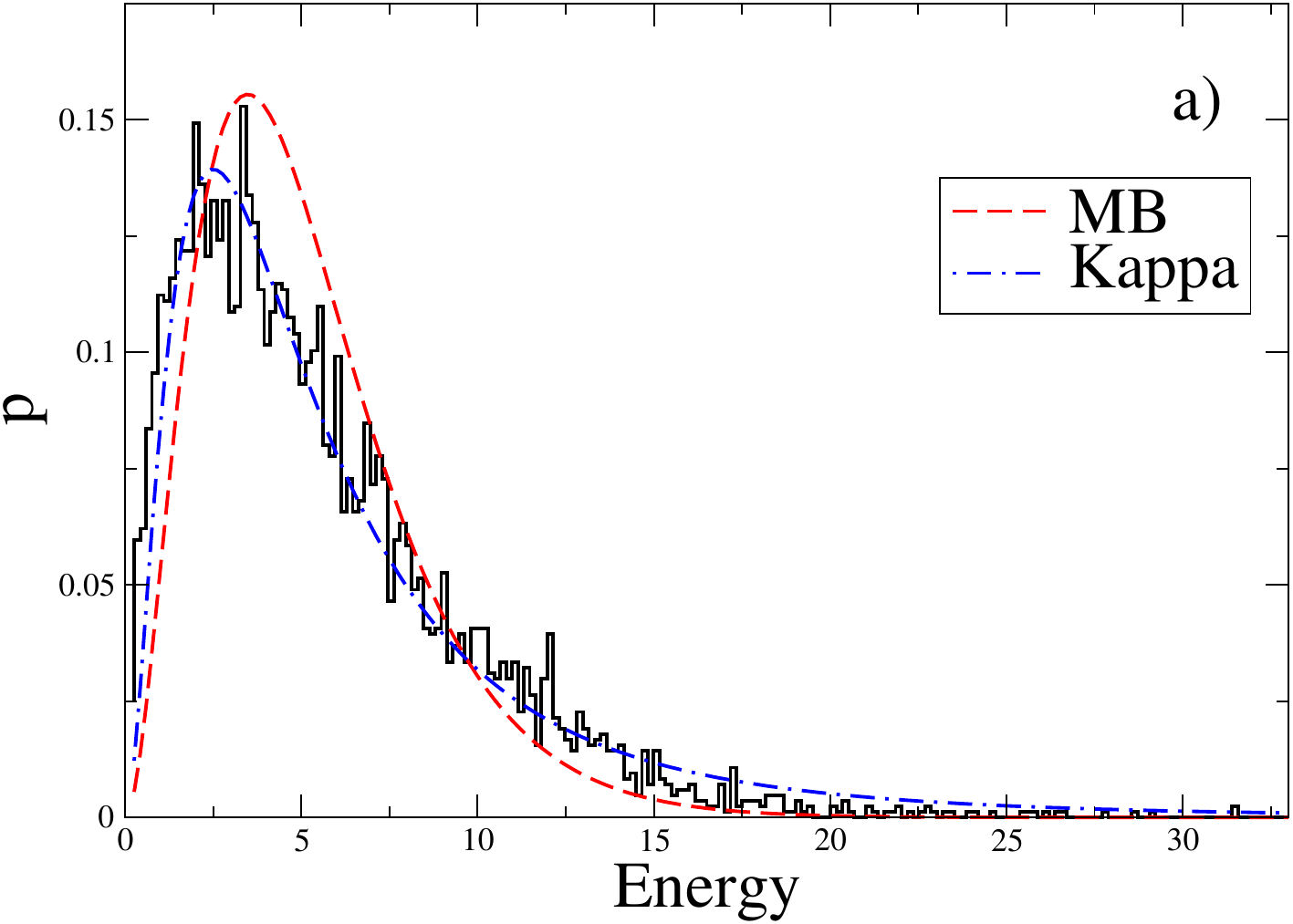}
\includegraphics[width=0.49\textwidth, clip=true]{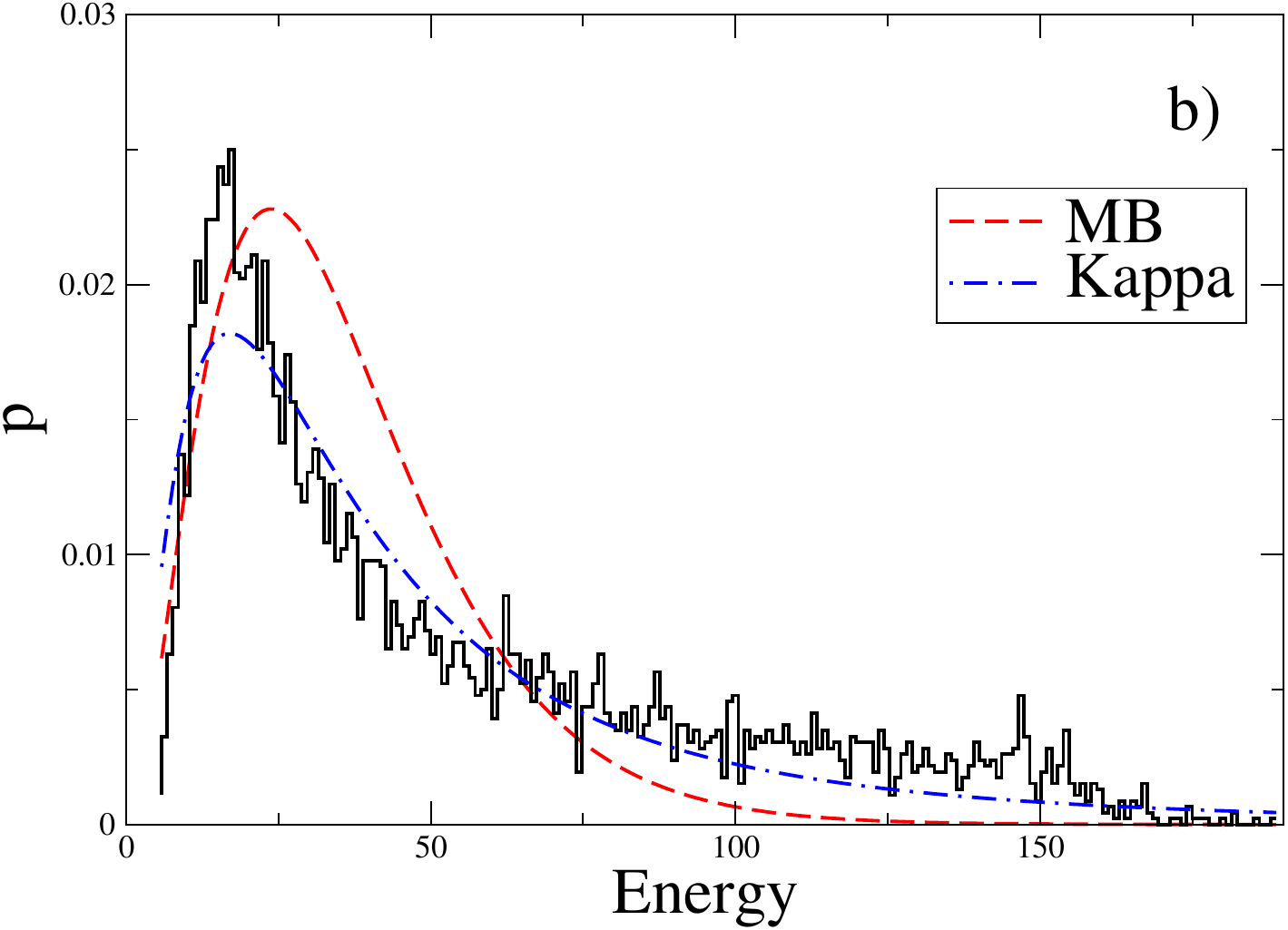}
\caption{Energy distribution for bath A  (at an inverse temperature of $\beta_A=2.0$ at the initial time) at $n_\mathrm{time}=200$ (a), and for bath B (at an inverse temperature of $\beta_B=0.2$ at initial time) 
at $n_\mathrm{time}=150$ (b), when $\xi_A$ and $\xi_B$ reach the minimum values in the data of Fig. 6b, respectively.  Also shown are the best fits with a Maxwell-Boltzmann 
distribution (red dashed) and a $\kappa$-distribution (blue dot-dashed). In the case of bath A, the best fit with a Maxwell-Boltzmann distribution yields 
$\beta_A=0.566$ with a $\tilde{\chi}^2=2.2 \times 10^{-4} $and a correlation coefficient of $0.953$, while 
the best fit with a $\kappa$-distribution yields $\beta_A=0.978$, $\kappa_A=5.51$, with 
a $\tilde{\chi}^2=5.7 \times 10^{-5}$, and a correlation coefficient of $0.984$.
In the case of bath B, we get best fits with a Maxwell-Boltzmann distribution for 
$\beta_B=0.084$ with a $\tilde{\chi}^2=1.2 \times 10^{-5}$ and a correlation coefficient of $0.902$.
The best fit with a $\kappa$-distribution yields $\beta_B=0.151$, $\kappa_B=4.55$, with a $\tilde{\chi}^2=3.8   \times 10^{-6}$ , and a correlation coefficient of $0.941$. Notice that both baths experience, at this selected $n_\mathrm{time}$, a doubling of their temperatures with respect to the initial values.} 
\label{Fig7}
\end{figure*}

\begin{figure*}[t]
\includegraphics[width=0.45\textwidth, clip=true]{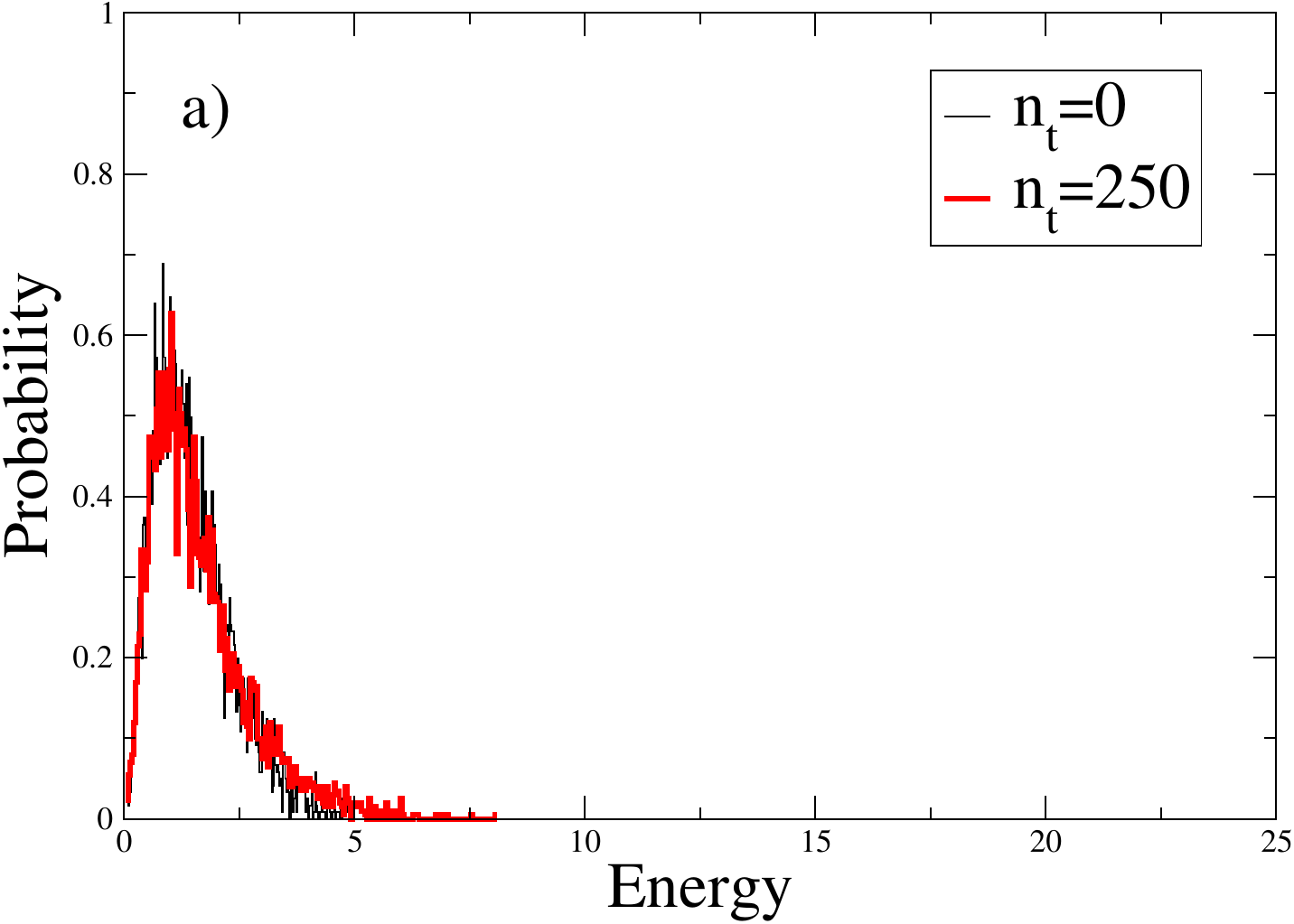}
\includegraphics[width=0.45\textwidth, clip=true]{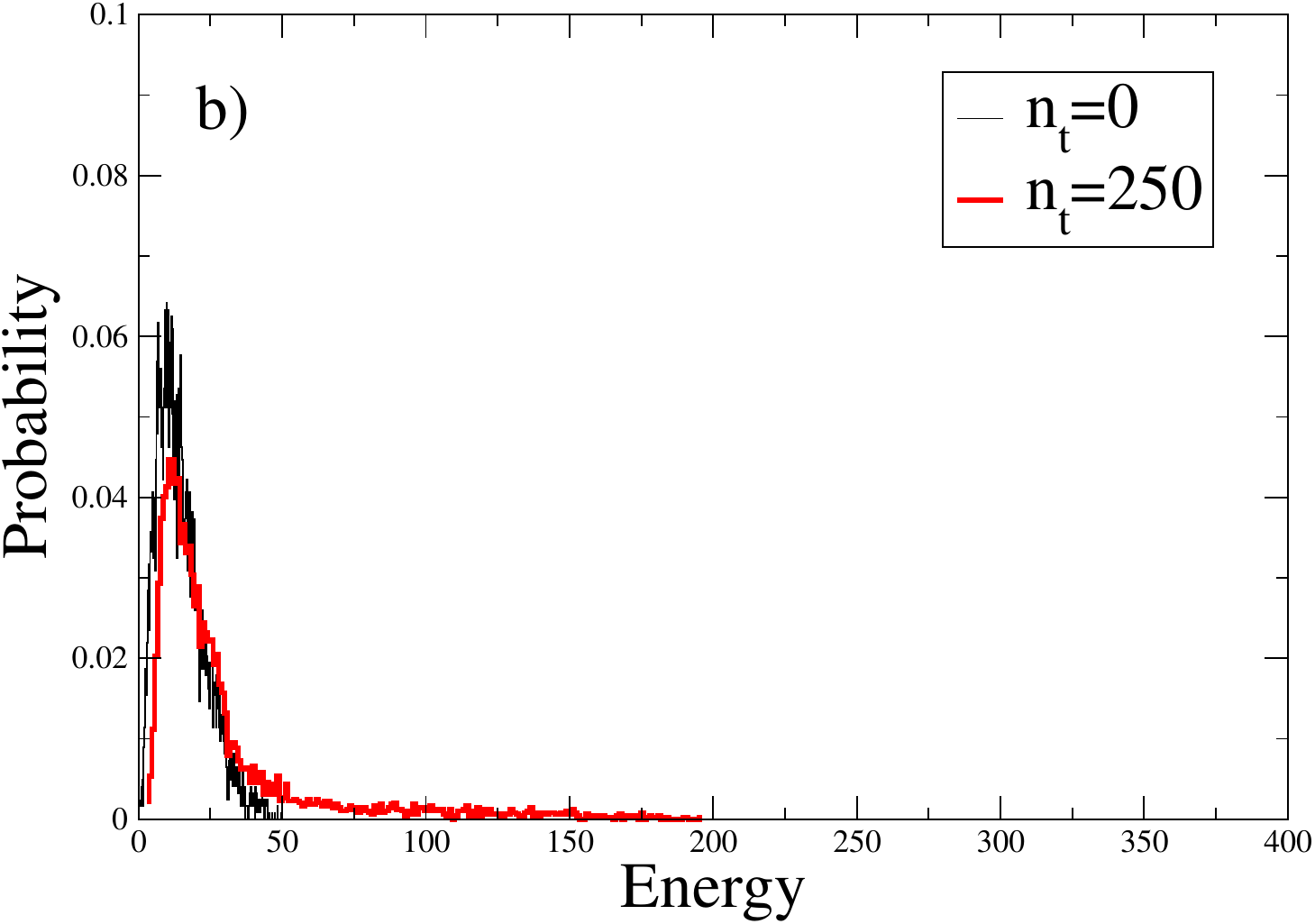}
\includegraphics[width=0.45\textwidth, clip=true]{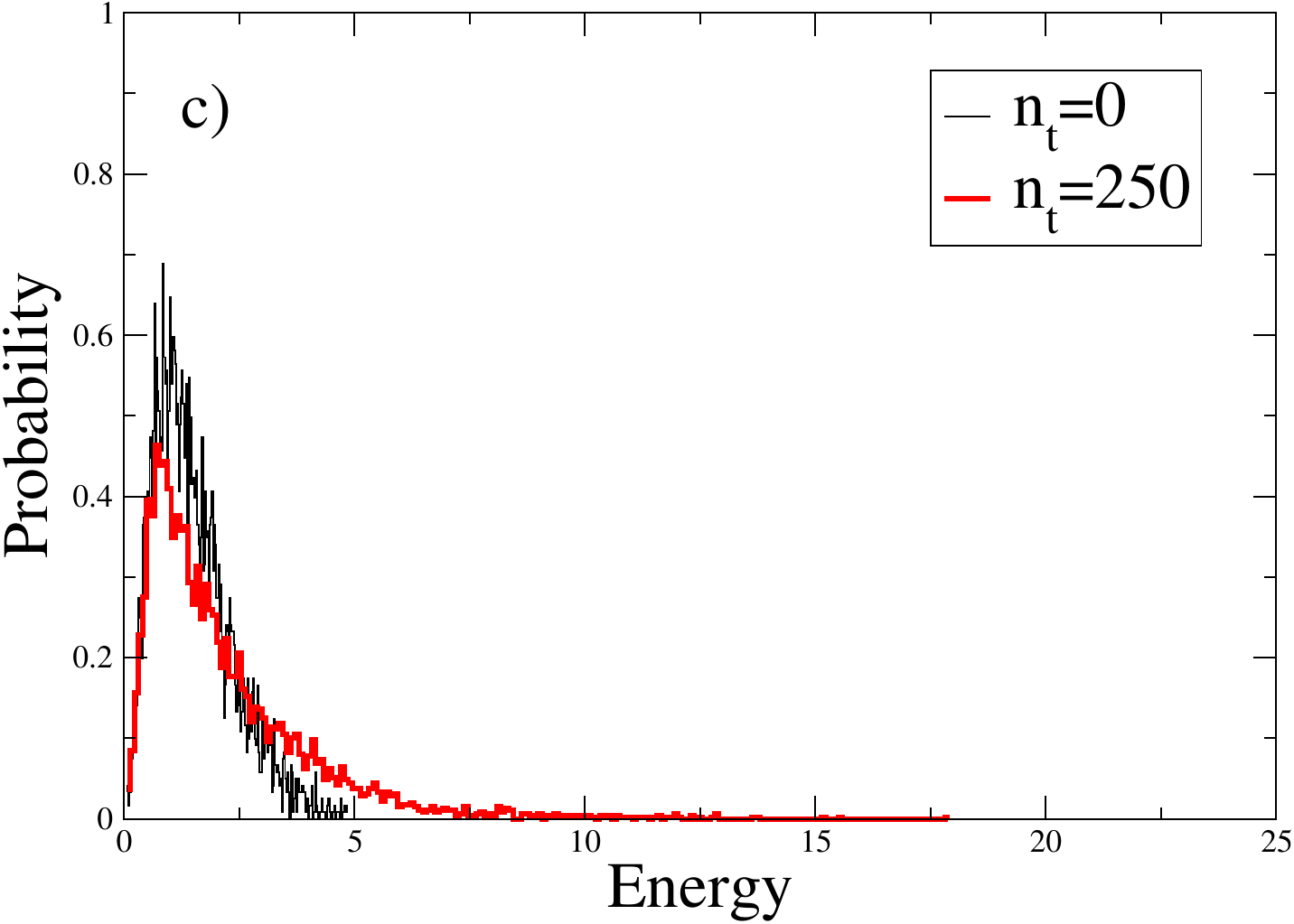}
\includegraphics[width=0.45\textwidth, clip=true]{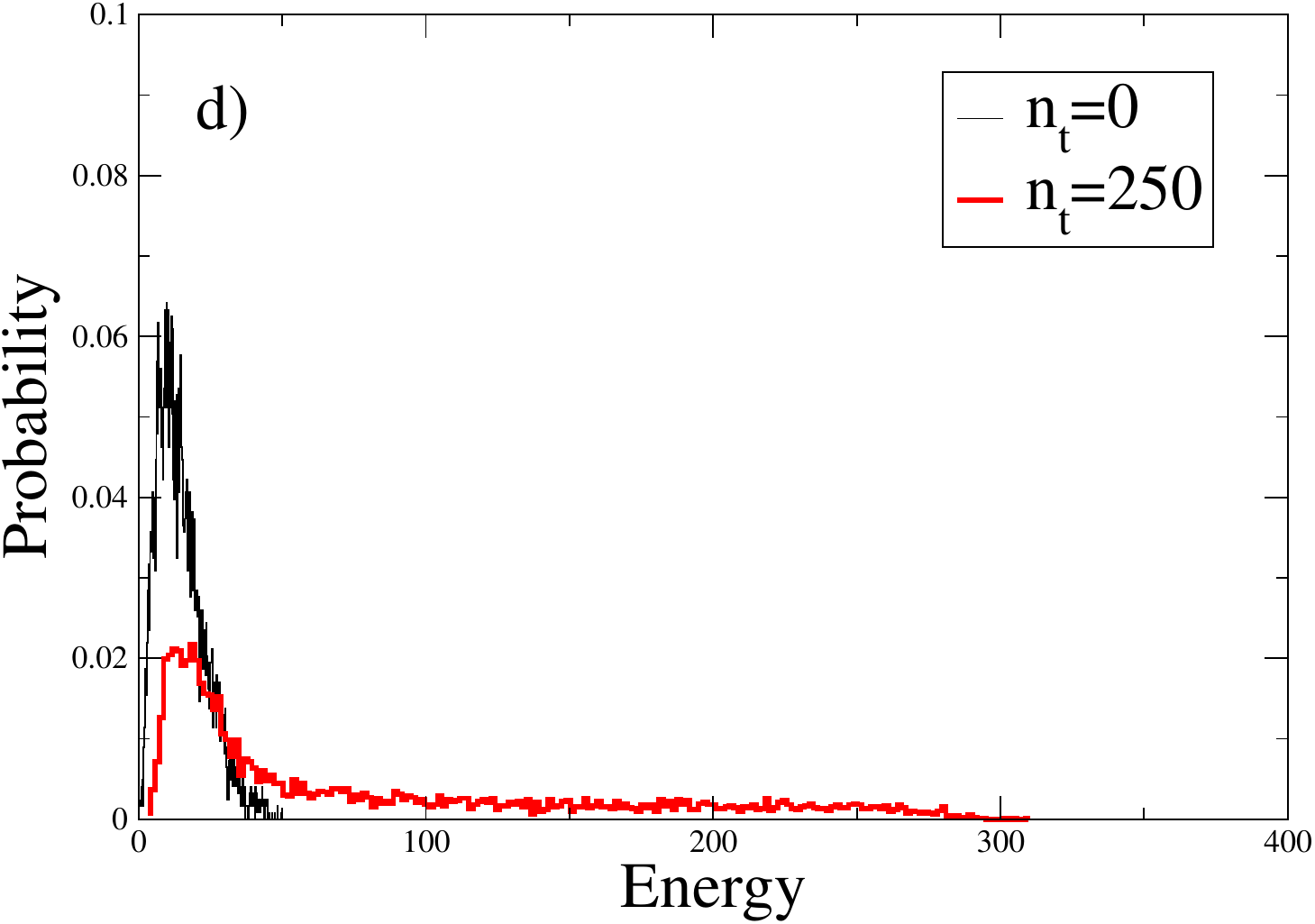}
\includegraphics[width=0.45\textwidth, clip=true]{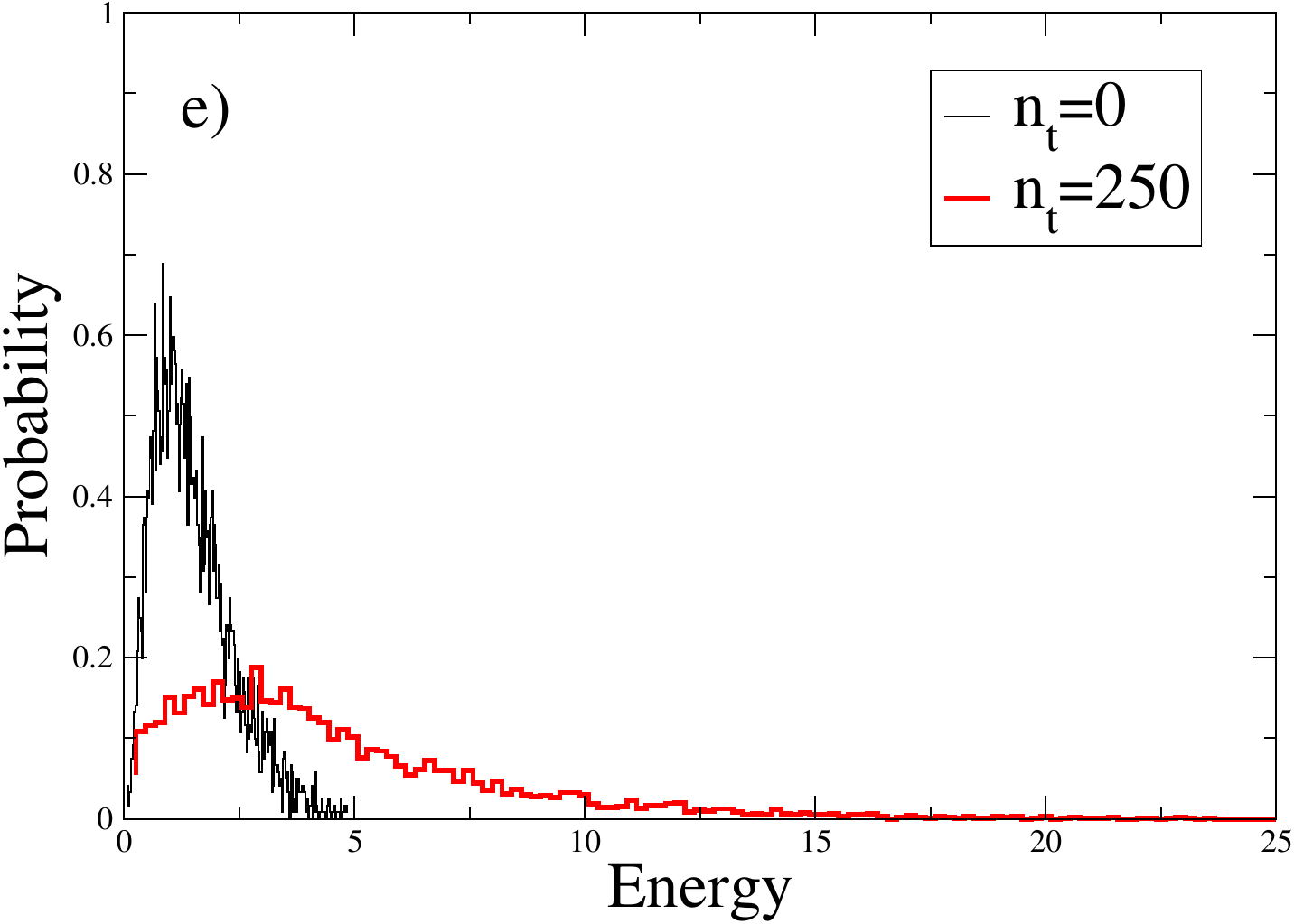}
\includegraphics[width=0.45\textwidth, clip=true]{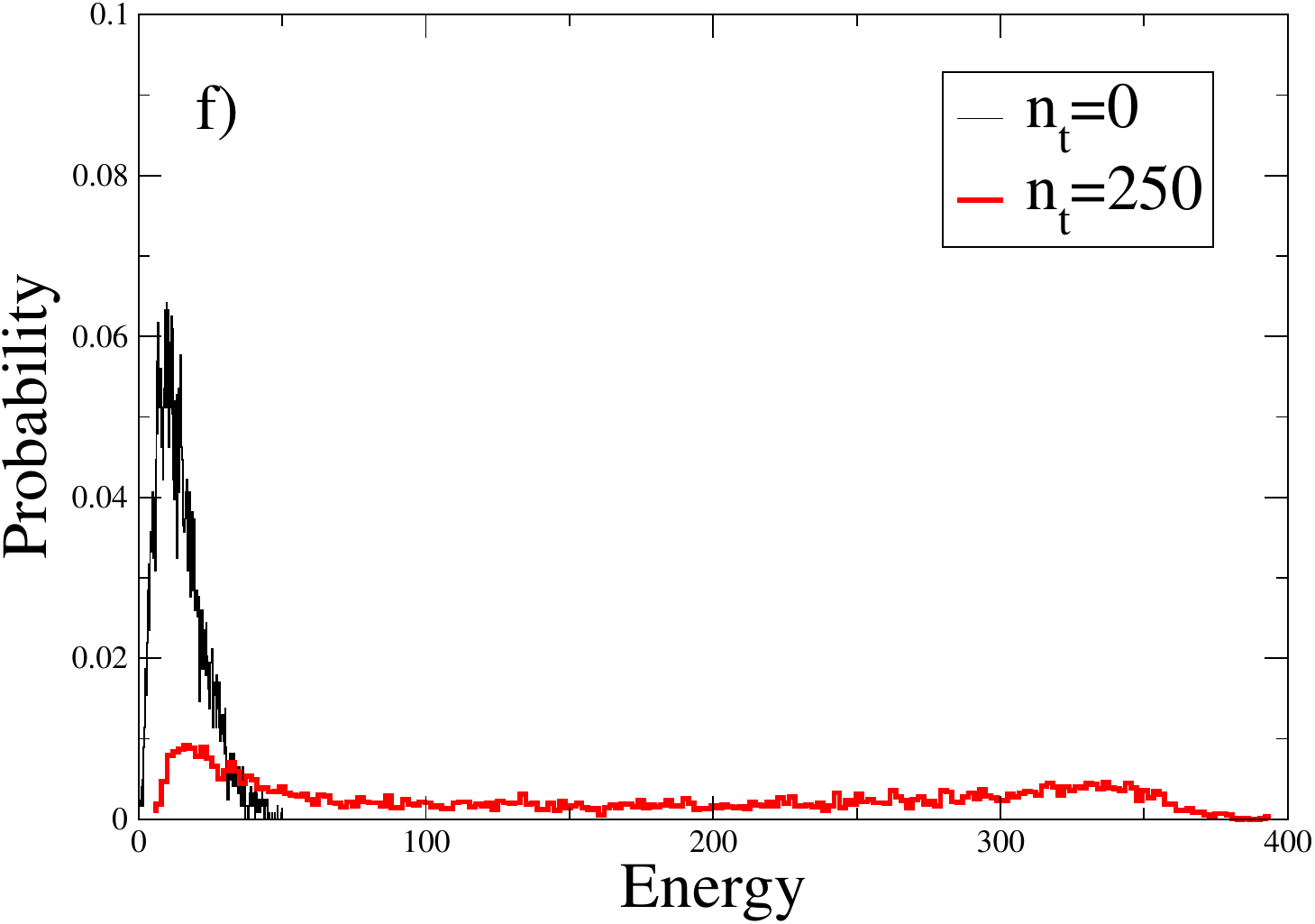}
\caption{Energy probability distributions evaluated after 250 time steps (thick red lines), starting from MB distributions (thin black lines), for values of $\gamma=0.1$ 
and $\lambda=1$, in the standard situation of equal mass and for the two baths made of $5000$ particles each, with frequency ratios 
$1:1:1$ in the top plots (a, b), $1:1:10$ in the middle plot (c, d), and $1:10:10$ in the bottom plots (e,f) for the cold bath (left plots) 
at $\beta_A=2$ and the hot bath (right plots)  at $\beta_B=0.2$.}
\label{Fig8}
\end{figure*}

\begin{figure*}[t]
\includegraphics[width=0.49\textwidth, clip=true]{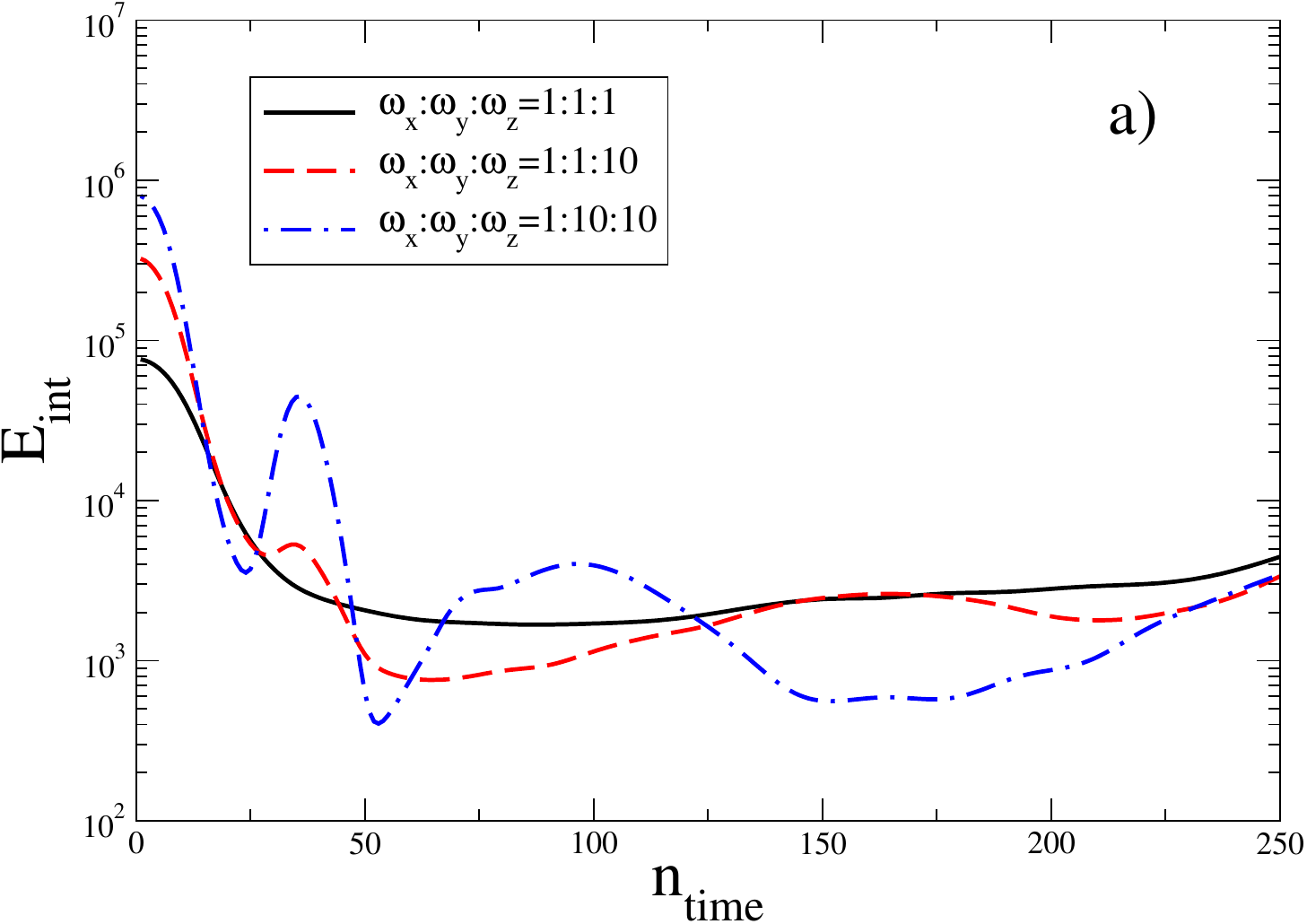}
\includegraphics[width=0.49\textwidth, clip=true]{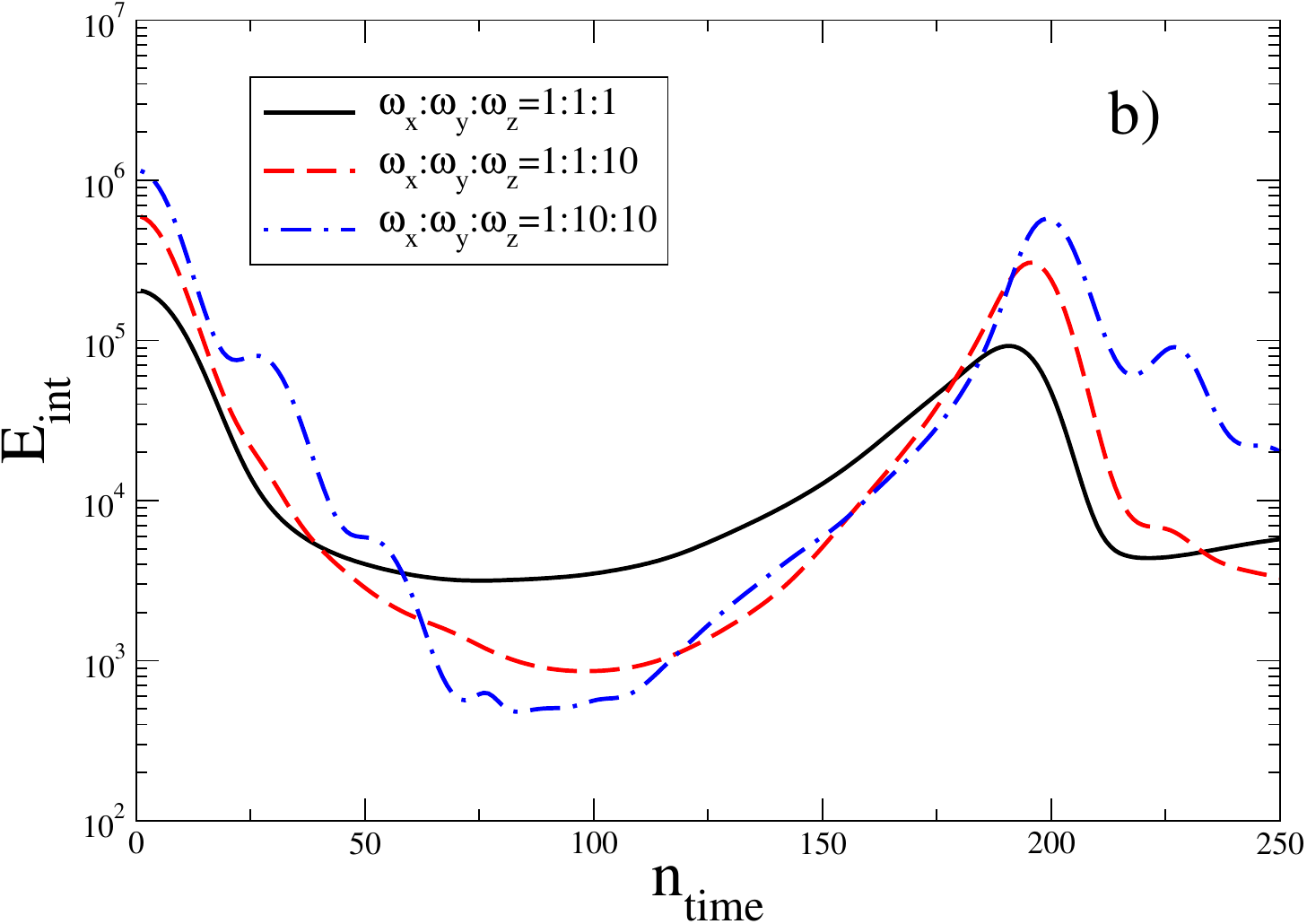}
\caption{Interaction energy versus time steps, for the same parameters of the case discussed in Fig.~\ref{Fig8}, and 
the frequency ratios indicated in the legends. The compression of the harmonic trap in the quasi-1D and quasi-2D cases 
results in larger interaction energies, corresponding to an higher effective interaction strength. Case (a) is for 
$\omega_A=\omega_B=1$, case (b) for $\omega_A=144/89, \omega_B=1$. }
\label{Fig9}
\end{figure*}

To explore the nature of these deviations from the MB expectation, we focused on 
the energy distributions coinciding with the times of maximal deviation. These occur at 200 and 150 time steps for bath A and B, respectively. We 
then contrasted the fit with a $\kappa$ energy distribution, characterized by the free parameters $\beta$ and $\kappa$, with the usual
MB energy distribution. The results are  shown in Fig.~\ref{Fig7}.  It is clear that the $\kappa$ energy distribution
 yields a better fit both in  terms of the position in energy of the peak as well as peak value, as well as in the high energy tail. 
 
Quantitatively, the fit results in a $\tilde{\chi}^2$ smaller by about a factor 3 with respect to the MB fit, and values of $\kappa_A=5.51, \kappa_B=4.55$. By using the corresponding values $\xi_A=0.63, 
\xi_B=0.62$ from Fig. \ref{Fig6}b we obtain,  either by solving Eq. \ref{XiDef} to get $\kappa$ from $\xi$, or simply by using Fig. \ref{Fig5}, $\kappa_A=5.31, \kappa_B=5.2$, rather consistent considering that the fit involves the entire energy distribution, not just the peak location.

Various energy distributions of the Tsallis or equivalent $\kappa$ form have been successfully utilized 
for fitting purposes in the data analysis of a broad range of phenomena \cite{Tsallis}. We are unaware of previous efforts where
these emerge in a many-body, first principle Hamiltonian model framework.
Of course, further work is needed to also understand distributions with $\xi>1$, as seen at later times in Fig.~\ref{Fig6}b, as well as the dependence of measures like $\xi$ on $\gamma$ and $\lambda$.  A preliminary consideration in this regard is that the largest deviations from $\xi=1$ appearing in the figure seems related to a fast change in the interaction energy involved in the dynamics, defined as 
\begin{equation}
E_{\mathrm{int}}=\gamma \sum_{m=1}^{N_A} \sum_{n=1}^{N_B} 
\exp{\left[-\frac{\left(Q_m-q_n\right)^2}{\lambda^2}\right]}.
\label{EqEint}
\end{equation}
This is clearly visible from the plots of $E_{\mathrm{int}}$ versus time steps appearing in Fig.~\ref{Fig6}.
It is reasonable to conjecture that a fast exchange of energy between the two baths can produce
strong deviations from the MB distribution, as we will explore in more detail in a future contribution.

Two final comments are in order. First, the presence of $\kappa$ energy distribution seems limited to very short times. With a period of bath A equal to $T_A=2\pi/\omega_B \simeq 3.88$ corresponding to about 388 times steps, the lowest value of $\kappa$ occurs around 200 times steps, {\it i.e.} at $\simeq 0.515~ T_A$. Similarly, the period of bath B being $T_B=2 \pi/\omega_B \simeq 6.28$, 
the lowest value of $\kappa$ occurs around 150 time steps, {\it i.e.} at $\simeq 0.24~ T_B$. 

Evidently, at early times the interaction is more effective in providing large and sudden injections of energy to the initial MB baths A/B, as the particles repeatedly move past each other in the harmonic potential. This results 
in a strongly non-adiabatic situation, as the particles  cannot quickly recover from the 
fast energy exchange. Second, the colder bath (A) prevails in cooling the hotter bath (B) in the central trapping region, and is instead subject to sudden heating in the peripheral region, an effect already discussed in \cite{Bala10}. The three-dimensional nature of the problem which reduces the likelihood of head-on collisions makes the recovery from this non-equilibrium effect of geometrical nature more difficult. 

\section{Short and long time thermalization dynamics in anisotropic traps}

\begin{figure*}[t]
\includegraphics[width=0.49\textwidth, clip=true]{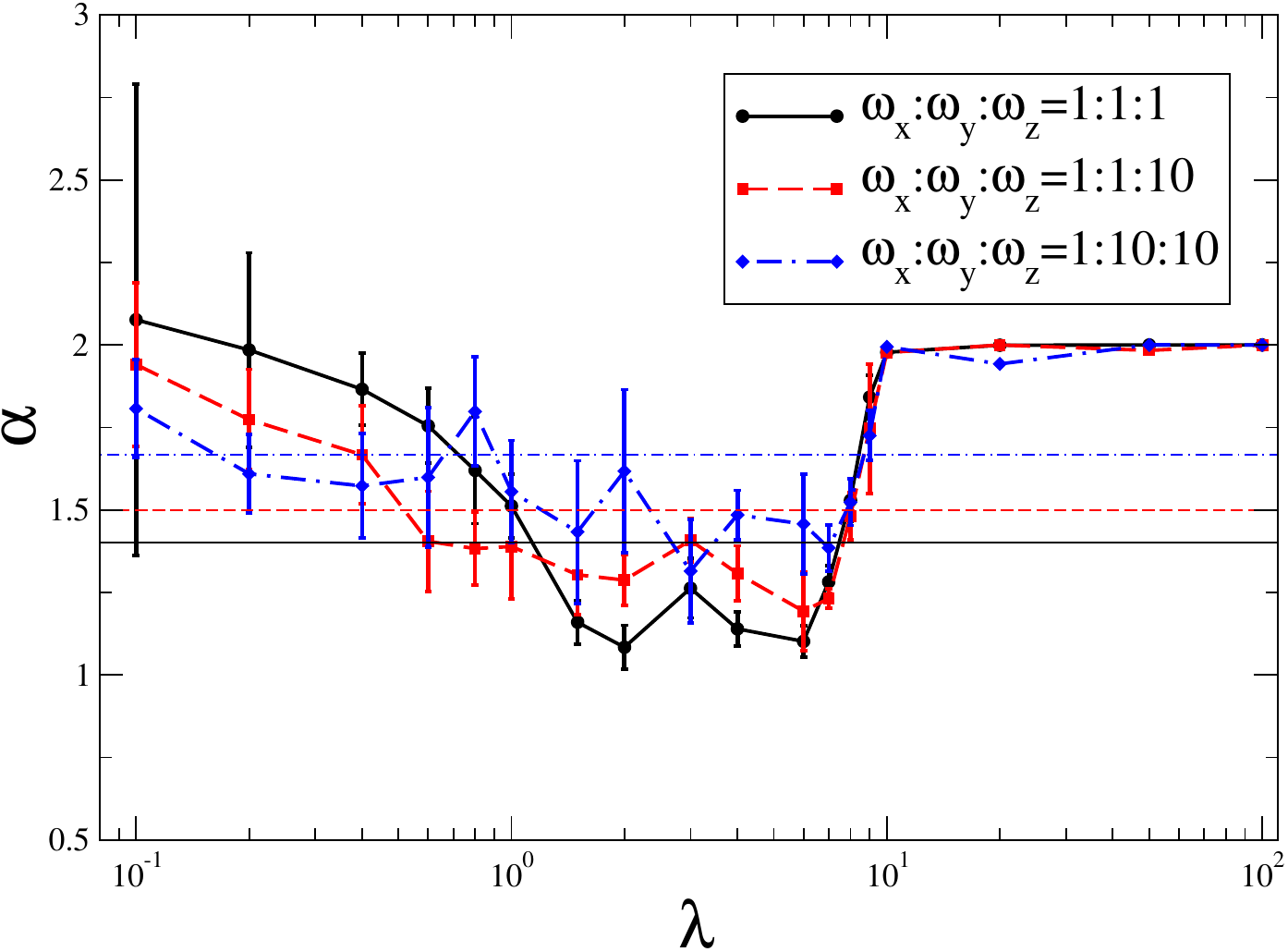}
\includegraphics[width=0.49\textwidth, clip=true]{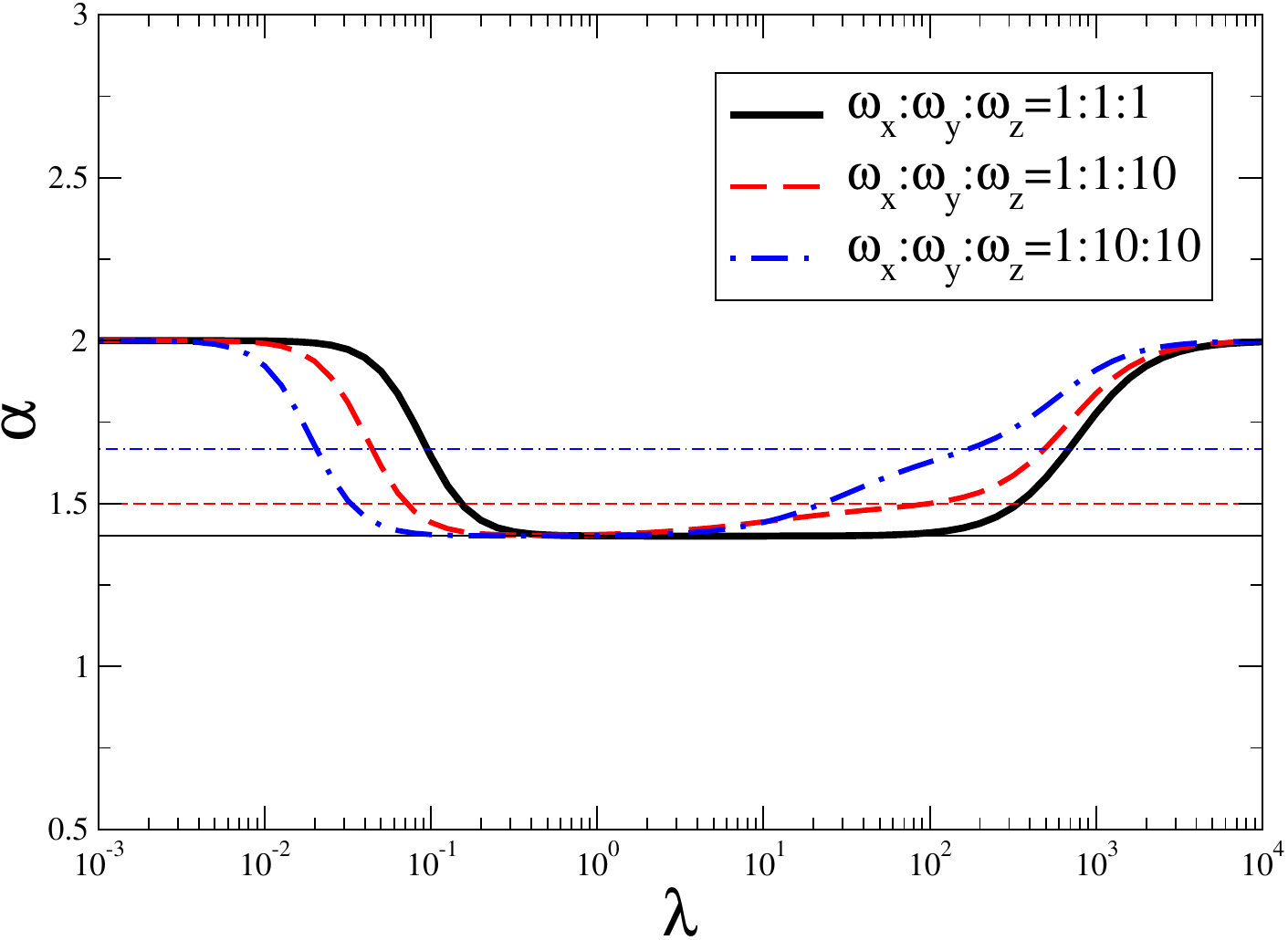}
\caption{Dependence of the scaling exponent $\alpha$ versus the interaction range $\lambda$ for three cases of trapping frequency ratios 
corresponding to an isotropic case, a quasi two-dimensional case, and a quasi one-dimensional case, with $\gamma=1$. The left plot is 
obtained by running the time evolution and collecting the final average interaction energy as described in the text (via Eq.~\ref{alpha}), the right plot by using the iterative equations Eqs. \ref{Iterative1}-\ref{Iterative2}. The horizontal lines corresponds to the expectations for $\alpha$ of 7/5, 3/2, and 5/3  for the isotropic, quasi-2D, and 
quasi-1D cases, respectively.}
\label{Fig10}
\end{figure*}

The analysis thus far has considered only isotropic traps. The addition of anisotropies add a further layer of considerations, as one can move from a full three dimensional dynamics to ones with lower effective dimensionality. This can drastically alter the frequency of the inter-species interactions, for instance due to the higher density of particles and the progressive increase of head-on collisions. The presence of anisotropy in trapping and effective temperatures may be beneficial in controlled fusion, as discussed 
in \cite{Nath,Kolmes,Li}. In this section we will investigate two distinct aspects of the problem. First, the transient features of thermalization for a general anisotropic trap, which reinforces the reasons why non-MB energy distributions are more difficult to generate in lower dimensions. Second, the stationary, long-time features such as the presence of anomalous scaling of the interaction energy with the number of particles. In Fig.~\ref{Fig8} we show an example by comparing the 
energy distributions at initial time and after 250 time steps for three cases, isotropic trapping with equal 
angular frequencies along the three axes, $\omega_{Ax}=\omega_{Ay}=\omega_{Az}=1:1:1$ (top plots), 
quasi-2D trapping, $\omega_{Ax}=\omega_{Ay}=\omega_{Az}=1:1:10$ (quasi 2D case), (middle plots), 
and quasi-1D trapping, $\omega_{Ax}=\omega_{Ay}=\omega_{Az}=1:10:10$ (bottom plots). 
In each row, the left column shows the distribution for the colder bath (A) while the right corresponds to
the hotter bath (B). All this is obtained by choosing an angular frequency ratio between the two baths, for each axis,
equal to two Fibonacci numbers,  $\omega_{Ai}/\omega_{Bi}=144/89$ with $i=x, y, z$, which should minimize recurrences and revivals for the two species.
A more prominent high-energy tail is visible in every one of the plots.
Any quantitative comparison between the three cases should account for the fact that as we increase the trapping frequencies in the 
quasi-2D and quasi-1D cases, the particles are progressively  more confined, with higher density and smaller average separation distance. 
This is evident in the comparison of the total interaction energy, as shown in Fig.~\ref{Fig9} ( these three frequency ratios are also used in Fig.~\ref{Fig10}). 
We note that the quasi-1D case has an initial interaction 
energy about one order of magnitude larger than the fully isotropic 3D case, with the quasi-2D being the intermediate case. 
The high-energy tails observed in the 
quasi-1D case should not be attributed to $\kappa$-distributions, and resemble more the bMB seen at 
large $\gamma$ in the strictly 1D case ~\cite{Bala10}.  We indeed observe the presence of a component resembling a MB energy 
distribution at low energy, and a `hump' at higher energies, clearly visible in the panel (f) of Fig.~\ref{Fig8} in the 300-350 energy range for the bath B. 
These can be viewed as a weighted sum of two MB distributions at different temperatures~\cite{Bala10}. As the $\kappa$ distribution can be constructed
by a weighted sum of a number of MB distributions at a range of temperatures, it is reasonable to speculate that the quasi-1D case reflects a precursor
to the $\kappa$ distribution show earlier in thermalization dynamics in the full 3D situation.

Another feature of the anisotropic cases is the larger time variability of the interaction energy, as more frequencies are involved in possible revivals. From the standpoint of a single bath, this implies a faster energy exchange with the external environment ({\it i.e.} the other bath) which creates a strong non-equilibrium situation.The revivals are only partially mitigated by using frequencies for the two baths with a ratio of two Fibonacci numbers, as seen from Fig.~\ref{Fig9}. By sampling the energy distributions of the colder bath at short times, we have 
explicitly verified that $\kappa$ energy distributions are a better fit than MB energy distributions, as manifestly visible in panels (c), and (e) of Fig.~\ref{Fig8}. This outcome is in line with the expectation that $\kappa$ energy distributions develop in the presence of systems in which the power is dissipated at a rate smaller than the incoming power, as discussed in \cite{Livadiotis1,Livadiotis2}. 
In this context, instances of time in which the interaction energy changes quickly should result in $\kappa$-distributed energies, motivation for further work on 
our model, and as seen from  Fig.~\ref{Fig9}, this is achieved more efficiently in anisotropic traps.

Finally we discuss the impact of anisotropic trapping on the anomalous exponent $\alpha$ for the scaling of the interaction energy $E_{\mathrm{int}}$ after thermalization upon the number of particles in the baths. As first shown in \cite{JauOnoSun1}, we found a power-law dependence that, in the case of baths with 
equal number of particles, $N_A=N_B=N$, can be written as $E_{\mathrm{int}} \propto N^\alpha$. We have numerically found that $\alpha$ is not uniformly equal to 1, as expected for an extensive situation  of short-range interactions, nor equal to 2 as expected for long-distance interactions. Numerically, we found that $\alpha$ ranges between these two extremes, depending on the parameters $\gamma$ \cite{JauOnoSun3} and $\lambda$ \cite{OnoSun2}. These results have been corroborated by analytic estimates obtained 
in the case of small coupling strength $\gamma$, as discussed in \cite{JauOnoSun3}, including the role of the dimensionality of the system.
The latter dependence is crucial especially in  suggesting a connection to the onset of turbulence in the physics of fluids and plasmas and the associated 
Kolmogorov scaling~\cite{Kolmogorov}. The dimensional arguments behind Kolmogorov scaling are scalar in nature, arising from assumptions of isotropy 
and resulting in effective one-dimensionality. In our dynamical situation, one-dimensionality is the situation favoring energy transfer due to the
absence of constraints on head-on collisions. In higher dimensions, angular momentum limits these, resulting in a slower rate of energy 
transfer, but we expect the differences to be quantitative. We explore this issue by comparing, in Table I and Fig.~\ref{Fig10}, fully 
isotropic trapping with frequencies selected to correspond to quasi 2D and
quasi 1D trapping situations. As in \cite{OnoSun2}, we have evolved the system until thermalization is achieved (in all cases this occurs within $10^4$ 
time steps). We have chosen to compare $E_{\mathrm{int}}$ after thermalization for the values of $N_1=200$ and $N_2=400$ particles, instead of the 
lower numbers as  in \cite{OnoSun2}, which results in smaller standard deviations on $\bar{E}_{\mathrm{int}}$. The latter is evaluated by 
considering the average of $E_{\mathrm{int}}(t)$ over the last $10^3$ time steps, and finally the exponent $\alpha$ 

\begin{equation}
\alpha=\log_2 \left[\frac{\bar{E}_{\mathrm{int}}(N_2)}{\bar{E}_{\mathrm{int}}(N_1)}\right],
\label{alpha}
\end{equation}
is obtained for various values of $\lambda$ and  $\gamma$. We expect weak dependence on $\gamma$,
at least for the relatively narrow range of values of $\gamma$ for which the anomalous scaling occurs. This was shown
in Figs. 4 and 5 in \cite{JauOnoSun3} and in the related discussion. This is further confirmed in Table I, where
we compare various cases of anisotropic trapping at constant $\lambda$ and two different values of $\gamma$.

The dependence on $\lambda$ is much richer as seen in the left plot of Fig. \ref{Fig10}. The scaling exponent $\alpha$ 
versus $\lambda$ is considered for a range of situations in three dimensions: an isotropic harmonic oscillator, a harmonic oscillator with trapping
 frequency ratio 1:1:10 (quasi-2D, `pancake' shaped), and a harmonic oscillator with trapping 
frequency ratio  1:10:10 (quasi 1D, `cigar'-shaped). In all cases, a decrease in $\alpha$ is seen for an intermediate 
range of $\lambda$ values. For $\lambda \ge 10$, corresponding to the long-range limit, we see convergence to $\alpha=2$. The same occurs for very small $\lambda$, although the presence of quasi-harmonic oscillations lead to large
fluctuations in $\alpha$. These are due to the 
systematic sloshing of the two clouds in the strong coupling regime (exhibited by a $\gamma/\lambda^2$ 
dependence in the classical equations of motion). Note also an inversion in the dependence of $\alpha$ which makes the 
quasi-1D case more anomalous with respect to the quasi 2D and isotropic cases. We attribute this to the increased effective 
$\bar{E}_{\mathrm{int}}$ in lower dimension, analogously to the effects on the energy distribution discussed in the previous section.
In the interval $1 \lesssim \lambda \lesssim 10$  the anomalous exponent values are in line with analytical estimates.

\begin{table}[t]
\begin{center}
\begin{tabular}{|c|c|c||c|c|c|}
       \hline \hline
 $\omega_y/\omega_x$  &  $\omega_z/\omega_x$    &     $\alpha(\gamma=1)$   & $\omega_y/\omega_x$  &  $\omega_z/\omega_x$  &  $\alpha(\gamma=2)$  \\ 
       \hline
1   &    1       &  2.046 $\pm$ 0.052  &   1 & 1   &  2.044  $\pm$ 0.033 \\
1   &    2       &  2.047 $\pm$ 0.089  &   1 & 1   &  2.003  $\pm$ 0.023 \\
1   &    4       &  2.038 $\pm$ 0.079  &   1 & 4   &  2.008  $\pm$ 0.024 \\
1   &    8       &  1.995 $\pm$ 0.047  &   1 & 8   &  1.922  $\pm$ 0.020 \\
1   &    16     &  2.002 $\pm$ 0.018  &   1 & 16 &  1.908  $\pm$ 0.029 \\
2   &    2       &  1.966 $\pm$ 0.026  &   2 & 2   &  1.951  $\pm$ 0.025 \\                          
2   &    4       &  1.973 $\pm$ 0.036  &   2 & 4   &  1.896  $\pm$ 0.015 \\
2   &    8       &  1.900 $\pm$ 0.011  &   2 & 8   &  1.923  $\pm$ 0.036 \\
2   &    16     &  1.874 $\pm$ 0.026  &   2 & 16 &  1.843  $\pm$ 0.059 \\
4   &    4       &  1.898 $\pm$ 0.023  &   4 & 4   &  1.848  $\pm$ 0.020 \\
4   &    8       &  1.847 $\pm$ 0.022  &   4 & 8   &  1.746  $\pm$ 0.056 \\
4   &    16     &  1.833 $\pm$ 0.057  &   4 & 16 &  1.582  $\pm$ 0.059 \\
8   &    8       &  1.744 $\pm$ 0.062  &   8 & 8   &  1.514  $\pm$ 0.066 \\
8   &    16     &  1.615 $\pm$ 0.069  &   8 & 16 &  1.396  $\pm$ 0.069 \\
16 &    16     &  1.420 $\pm$ 0.092 & 16 & 16 &  1.235  $\pm$ 0.069 \\
\hline
\hline
\end{tabular}
\caption{Values of the exponent $\alpha$ for a variety of anisotropic trapping situations, for two different values 
of the interaction strength $\gamma$, interpolating from the fully isotropic case to the intermediate case of an effective
two-dimensional trapping ($\omega_x:\omega_y:\omega_z=1:16:16$), until an effective one-dimensional trapping 
($\omega_x:\omega_y:\omega_z=1:1:16$). In these runs we have $\lambda=0.1$, $\beta_A=2.0$, $\beta_B=0.2$,
and the absolute value of the trapping angular frequencies in the $x$-direction are $\omega_A=144/89$, $\omega_B=1.0$.}
\end{center}
\label{tab:tableexponent}
\end{table}

In an earlier publication \cite{JauOnoSun3}, we had numerically shown power-law scaling of the interaction
energy with particle number, where the exponent allowed us to distinguish different regimes
of thermalization as a function of the model parameters, $\gamma$ and $\lambda$. We also 
constructed simple thermodynamic arguments to analytically recover the exponents in a
regime that was analogous to turbulent fluid or Kolmogorov mixing. This analysis can be
generalized to a more generic anisotropic case.

The Hamiltonian in Eq.~(\ref{Hamilton}) may be seen as the sum of two terms, the free Hamiltonian of two sets of harmonic oscillators (without any dependence on $\gamma$ and $\lambda$) and their interaction Hamiltonian, $H=H_0+\gamma I$, with $I$ a dimensionless quantity.
The corresponding partition function and the expectation value of energy at thermal equilibrium at inverse temperature $\beta$ are, respectively,

\begin{equation} 
Z=\int d\vec{p}\, d\vec{r} \, \exp{\left(-\beta H_0 - \beta \gamma I\right)},
\label{Partition}
\end{equation} 

\begin{eqnarray} 
\langle E \rangle &=&-\frac{1}{Z} \frac{\partial Z}{\partial \beta } \\
&=&
\frac{1}{Z} \int \int d\vec{p}\, d\vec{r} \left(H_0 + \gamma I\right) 
\exp(-\beta H_0 -\beta \gamma I). \nonumber
\label{PartitionEnergy} 
\end{eqnarray} 

The ensemble average of the interaction energy of any pair of interacting particles can be obtained by 
assuming that the particles in the two components of the mixture satisfy the Maxwell-Boltzmann distribution at long times. 
We expand the expression of the energy in terms of the coupling strength $\gamma$, obtaining, in the case of a balanced mixture
($N_A=N_B=N$)

\begin{eqnarray} 
\langle E\rangle & = &\frac{6N}{\beta}+ N^2 \gamma \lambda^6 \left(\frac{m_A m_B}{4}\right)^{3/2}\beta^3 
\prod_{i=x,y,z}\omega_{Ai} \omega_{Bi} \times \nonumber \\
&& \left[\left(1+\frac{\beta \lambda^2 m_A \omega_{Ai}^2}{2}\right)\left(1+\frac{\beta \lambda^2 m_B \omega_{Bi}^2}{2}\right)-1\right]^{-1/2}. 
\label{EnergyApprox1} 
\end{eqnarray}
This becomes simpler if expressed in terms of the ratio between the interaction range $\lambda$ and the thermal 
lengths of the two species in each direction $i=x, y, z$, {\it i.e.}
\begin{eqnarray}
\xi_{Ai}&=&\sqrt{2/(\beta m_A \omega_{Ai}^2)}, \\
\xi_{Bi}&=&\sqrt{2/(\beta m_B \omega_{Bi}^2)},
\label{Thermallength}
\end{eqnarray}
using the inverse temperatures $\beta$. The resulting expression is
\begin{eqnarray} 
&&\langle E\rangle = \frac{6N}{\beta}+ N^2 \gamma  \prod_i \frac{\lambda}{\xi_{Ai}} \frac{\lambda}{\xi_{Bi}} \\
&& \times\left\{\left[1+\left(\frac{\lambda}{\xi_{Ai}}\right)^2\right]
\left[1+\left(\frac{\lambda}{\xi_{Bi}}\right)^2\right]-1\right\}^{-1/2}. \nonumber
\label{EnergyApprox2} 
\end{eqnarray}

Thus the average interaction energy $\langle E_{\mathrm{int}} \rangle$, for $\gamma$ consistent with the perturbative approximation,
depends on the amplitudes of the directional thermal lengths relative to the interaction range $\lambda$.  This fully
captures the interplay between scaling exponent $\alpha$ and the trapping anisotropy. 
Based on the comparison of these length scales, various cases may be discussed, and scaling exponents can be derived 
by requiring homogeneity for the dependence of the total energy upon the number of particles, by setting a relationship 
between the equilibrium temperature and the number of particles~\cite{JauOnoSun3}. 
For instance, if $\lambda/\xi_{Ai}, \lambda/\xi_{Bi} \gg 1$ in all directions and for both baths, then the term in the 
multiproduct of Eq.~(\ref{EnergyApprox2}) is approximately unity, and then the interaction energy becomes 
independent on both $\lambda$ and the inverse temperature $\beta$. 
\begin{equation}
\langle E_{\mathrm{int}} \rangle \simeq  N^2 \gamma,
\end{equation}  
which allows homogeneity with the equipartition term if we choose $\beta \propto N^{\tau}$ with $\tau=-1$, corresponding to
$\langle E\rangle \propto N^2$.

In the opposite, extreme case where instead $\lambda/\xi_{Ai}, \lambda/\xi_{Bi} \ll 1$,
the terms in the product are approximated as
\begin{eqnarray}
\prod_i & & \frac{\lambda}{\xi_{Ai}} \frac{\lambda}{\xi_{Bi}} 
\left[\left(\frac{\lambda}{\xi_{Ai}}\right)^2 +\left(\frac{\lambda}{\xi_{Bi}}\right)^2\right]^{-1/2} \nonumber \\
\sim \prod_i & & \beta^{1/2} \propto \beta^{3/2}.
\end{eqnarray}  
This implies 
\begin{equation}
\langle E_{\mathrm{int}} \rangle \propto \gamma N^2 \beta^{3/2},
\end{equation}  
and, again setting $\beta \sim N^{\tau}$, we get homogeneity if $1-\tau=2+3\tau/2$, {\it i.e.} $\tau=-2/5$, and
$\langle E\rangle \sim N^{7/5}$. By repeating the analysis in the intermediate case of
$\lambda/\xi_{Ai}, \lambda/\xi_{Bi} \ll 1$ only valid for two of the three dimensions, we get
\begin{equation}
\langle E_{\mathrm{int}} \rangle \propto \gamma N^2 \beta,
\end{equation}
which leads to $1-\tau=2+\tau$, {\it i.e.} $\tau=-1/2$ and $\langle E\rangle \sim N^{3/2}$.
Finally, if $\lambda/\xi_{Ai}, \lambda/\xi_{Bi} \ll 1$ only for one of the three dimensions, we get
\begin{equation}
\langle E_{\mathrm{int}} \rangle \propto \gamma N^2 \beta^{1/2},
\end{equation}
which leads to $1-\tau=2+\tau/2$, {\it i.e.} $\tau=-2/3$ and $\langle E\rangle \sim N^{5/3}$.

We therefore recover the Kolmogorov scaling exponents of 5/3, 3/2, and 7/5 based on the effective dimensionality
of the systems, as already discussed in~\cite{JauOnoSun3}.
Of course more complicated situations are expected if the two systems are unbalanced
in mass and oscillation frequency, differ in their particle number, or if crossed anisotropies (in which the 
trapping frequencies ratios for the two systems differ for the three directions, creating for instance a pancake geometry 
for one system, while the other is in a cigar-shaped geometry) are present.

Analytical estimates for the dependence of $\alpha$ upon the interaction range $\lambda$ can also be obtained 
using a recursive, finite difference equation. 
Even far from global equilibrium, we can define the total energy as sum of terms containing two distinct inverse temperatures 

\begin{eqnarray} 
\label{EnergyAppr} 
&&\langle E\rangle = 6 \left(\frac{N_A}{\beta_A}+\frac{N_B}{\beta_B}\right)+ N^2 \gamma \times \\ 
&&\prod_i \frac{\lambda^2}{\xi_{Ai}\xi_{Bi}} \left\{\left[1+\left(\frac{\lambda}{\xi_{Ai}}\right)^2\right]
\left[1+\left(\frac{\lambda}{\xi_{Bi}}\right)^2\right]-1\right\}^{-1/2}, \nonumber
\end{eqnarray}
where  $\xi_{Ai}$ and $\xi_{Bi}$ have been redefined to include two distinct $\beta_A$ and $\beta_B$ which
results in
\begin{equation} 
\langle E\rangle = 6 \frac{N_A}{\beta_A}+ 
\frac{N_A \langle E_{\mathrm{int}} \rangle}{N_A+N_B}  + 
6 \frac{N_B}{\beta_B} +
\frac{N_B\langle E_{\mathrm{int}} \rangle}{N_A+N_B}.
\label{Energy4} 
\end{equation}

We assume that the thermalization processes are such that the interaction energy is quickly absorbed 
into each system independently, and according to the heat capacity of each system. This assumption 
corresponds to faster intraspecies thermalization as compared with interspecies thermalization, 
which may be invalid in the regime of very strong coupling $\gamma >> 1$.
We introduce two new inverse temperatures $\beta_A'$ and $\beta_B'$ such that the interaction energy 
is reabsorbed into their definition

\begin{eqnarray} 
& & \frac{N_A}{\beta_A}+ \frac{N_A  \langle E_{\mathrm{int}} \rangle}{6(N_A+N_B)}  = \frac{N_A}{\beta_A'}, \\
& & \frac{N_B}{\beta_B}+ \frac{N_B \langle E_{\mathrm{int}} \rangle}{6(N_A+N_B)}  = \frac{N_B}{\beta_B'}.
\end{eqnarray}

This corresponds to iterative equations of the form
\begin{eqnarray} 
\frac{1}{\beta_{A(n+1)}} &=& \frac{1}{\beta_{A(n)}} + \frac{\langle E_{\mathrm{int}} \rangle_{(n)}}{6(N_A+N_B)}, 
\label{Iterative1}\\
\frac{1}{\beta_{B(n+1)}} &=& \frac{1}{\beta_{B(n)}} + \frac{\langle E_{\mathrm{int}} \rangle_{(n)}}{6(N_A+N_B)}. 
\label{Iterative2}
\end{eqnarray}
where $\langle E_{\mathrm{int}} \rangle_{(n)}$ depends upon $\beta_A$ and $\beta_B$ as in the second term in the right  hand side of Eq.~\ref{EnergyAppr}.
Iterating these equations may lead to fixed points corresponding to the equilibrium situation with $\beta_{A{(n+1)}}=\beta_{B{(n+1})}={\mathrm{const}}$ as $n \rightarrow \infty$.
In the Caldeira-Leggett limit, when $\lambda/\xi_{Ai}, \lambda/\xi_{Bi} \gg 1$, we have already shown that the interaction energy does not depend upon 
temperature, so the fixed point seems to occur at $1/\beta_A=1/\beta_B=\infty$. 
This situation of diverging temperature for both systems is obviously  
incompatible with the constraints of a Hamiltonian system with constant and finite energy. In other words, this confirms from another perspective that 
in a long-range system with just two frequencies there is no thermalization at any timescale~\cite{Smith}. The opposite situation of 
$\lambda/\xi_{Ai}, \lambda/\xi_{Bi} \ll 1$ is more interesting, since in this case, using the approximate form of the interaction energy discussed above, we have

\begin{eqnarray} 
\frac{1}{\beta_{A(n+1)}} &=& \frac{1}{\beta_{A(n)}} + \frac{\gamma N_A}{2} \beta_{A(n)}^{3/2}, \\
\frac{1}{\beta_{B(n+1)}} &=& \frac{1}{\beta_{B(n)}} + \frac{\gamma N_B}{2} \beta_{B(n)}^{3/2}.
\end{eqnarray}

It is easy to check that in this case there is a fixed point corresponding to an inverse temperature $\beta_{\mathrm{eq}}=[4/(3 \gamma N)]^{2/5}$. 
This generates in turn the correct scaling for the interaction energy with exponent $7/5$ as expected for the fully 3D case. Notice also that the 
inverse temperature at equilibrium is inversely proportional to $\gamma$ and $N$, and that due to its dependence on $N$ it loses its intensive character.

In the right plot of Fig.~\ref{Fig10} we show the scaling exponent versus $\lambda$ generated by evaluating the dynamics described above, and 
in the left plot we show the same dependence for the scaling exponent obtained by computer simulations for two different values of number of particles, and assuming a power-law scaling, as described in detail in \cite{OnoSun2}, for various degrees of anisotropy. 
Considering the crude approximations in deriving the recursive equations, there is general agreement with the numerical evaluations of $\alpha$ versus 
$\lambda$.  In particular, it is worth noticing that there is a similar crossover of the quasi-1D and quasi-2D cases at 
an intermediate value of $\lambda$, leading to less anomalous behavior with decreasing $\lambda$. The expected scaling exponents of 5/3,  3/2, and 7/5 in a Kolmogorov-like approach are only achieved for a restricted range of $\lambda$ values, with the deviations from these exponents more prominent for the quasi-2D and quasi-3D cases.

\section{Conclusions}

We have considered thermalization processes for two baths made of harmonic oscillators with special emphasis on features which appear in the three-dimensional case. By focusing on the location of the peak of the energy distribution, we have identified a transient behavior in which the two baths have energies deviating from the Maxwell-Boltzmann distribution otherwise present at initial and at long times. 
These deviations are in line with what is expected for $\kappa$ or $q$ distributions in the case of long-range interactions, a general hypothesis not dependent on the
explicit form of the interaction term and analogous to the emergence of 
MB distributions as discussed in \cite{Khinchin}.  The presence of $\kappa$ distributions at early times in a thermalization process may play a relevant role in nuclear fusion as a mean to enhancing the reactivity  \cite{Majumdar,Onofrio,Xie1,Xie2,Kong,Squarer,Ourabah}. 

We also discussed the impact of trapping anisotropies on the dynamics of thermalization, including deviations from intensivity of quantities expected in the usual thermodynamic setting. Anomalous scaling in this model, initiated in \cite{JauOnoSun3}, is reminiscent of Kolmogorov scaling for turbulent phenomena \cite{Kolmogorov}, and this makes a connection with the onset of turbulence in lower dimensional plasmas \cite{Antar,Mason}. In a more practical context, the presence of deviations from Maxwell-Boltzmann energy distributions, possibly with a more robust component of high-energy particles, is beneficial for the suppression of turbulence in confined plasmas, of interest in nuclear fusion \cite{Wilkie,Garcia1,Mazzi,Garcia2,Citrin}. Moreover, this contribution is in line with the 
proposed study of critical properties of plasmas by use of analogies to the theory of fluids, discussed in detail in \cite{Montgomery}, using Hamiltonian systems \cite{Morrison1,Morrison2,Morrison3}.

\end{document}